# Ultrashort echo time and zero echo time MR imaging and their applications at high magnetic fields: A literature survey

Soham S. More[1], Xiaoliang Zhang[1,2]

[1]Department of Biomedical Engineering, School of Engineering and Applied Sciences, and Jacobs School of Medicine and Biomedical Sciences, State University of New York at Buffalo, Buffalo, NY 14260, USA
[2]Department of Electrical Engineering, School of Engineering and Applied Sciences, State University of New York at Buffalo, Buffalo, NY 14260, USA

## Abstract

Ultrashort Echo Time(UTE) and Zero Echo Time (ZTE) imaging sequences have been developed to detect short T2 relaxation signals emanating from regions/tissues that conventional MRI methods fail to capture. Due to the high dipole-dipole interactions present in solid and semi-solid tissues, the signal decay is very fast, typically less than 1 millisecond. The echo time generated by conventional imaging methods is insufficiently short, resulting in void signals from these short-T2 species. The application of these UTE and ZTE techniques allows for imaging of solid and semi-solid tissues, making the conventionally invisible visible, which can significantly impact clinical imaging.

High and ultra-high field strength (UHF) MRI offers a critical advantage in enhancing the sensitivity and specificity of MR imaging. When combined with UTE and ZTE sequences, it enables the recovery of signals from void areas and markedly improves signal quality. To bolster this approach, intensive efforts have been made to collect supplementary data from various research tools to further validate the technique while addressing its limitations.

Studies demonstrate that the integration of UTE and ZTE sequences, along with innovative high-field imaging techniques in transmit/receive hardware and software, especially the exploration of new trajectory approaches, plays a pivotal role in enhancing and accurately depicting musculoskeletal, neural, lung, and dental MR imaging.

## Introduction

Magnetic Resonance Imaging (MRI)[1, 2] has firmly established itself as a promising diagnostic method in clinical imaging[3]. Recent advancements in clinical scanners and their increased accessibility have ignited widespread interest in high-field (3-5 Tesla) and ultra-high-field MRI (7 Tesla and beyond)[4-33]. Notably, one of the compelling advantages of higher field strengths is the substantial improvement in Signal-to-Noise Ratio (SNR), which, in in-vivo systems, increases linearly (and often nearly quadratically within certain field strength ranges)[34-44]. This enhanced sensitivity offers a distinct advantage over similar MRI conducted at lower field strengths[45], albeit accompanied by numerous technical challenges, particularly in the realm of radiofrequency hardware at ultrahigh field technology[46-81].

Within the human body, various tissues exhibit short T2 relaxation times, spanning from mere microseconds to several hundreds of microseconds[45, 82, 83]. Furthermore, certain tissues and associated structures possess extremely short T2 relaxation times (<1ms), rendering them undetectable through conventional means primarily due to limitations in achieving a minimum echo time. To address the imaging of these components, Ultra-short Echo Time (UTE) and Zero Echo Time (ZTE) MRI techniques were developed[84, 85]. Both methods employ specialized acquisition and reconstruction techniques tailored to visualize solid and semi-solid structures, including tendons, cortical bone, lung parenchyma, myelin components, ligaments, and related tissues[86]. A comprehensive list of T2/T2* values for human organs and tissues is provided in Table 1.

Transverse relaxation time depends on both the stiffness or hardness of tissues and the magnetic field strength. As stiffness increases, T2 relaxation time decreases. Similarly, with an increase in magnetic field strength, T2 relaxation time decreases. Given that most solid and semi-solid tissues are associated with the musculoskeletal system, neurological system, and the lungs, our focus will be on investigating the outcomes of applying these techniques to these organ systems.

In this review, we will delve into the applications of UTE and ZTE techniques at varying magnetic field strengths, with a particular emphasis on ultrahigh fields. We will also provide a concise summary of the recent developments in these techniques. Furthermore, we will explore various hardware considerations necessary for designing and optimizing functionality to achieve the desired outcomes.

Table 1: Tissues and their T2/T2* relaxation times at different field strengths of 1.5T, 3.0T and 7.0T, respectively. The stiffness of the tissue was highly related to its T2 relaxation time, with decreasing T2 time for increasing stiffness. T2 characteristics also had an inverse relation with field strengths, with decreasing values for increasing magnetic field strength.

| | **TISSUE** | **1.5T** | **3T** | **7T** |
|---|---|---|---|---|
| **SOLID AND SEMI-SOLID TISSUES** | Cortical bone | 0.45ms[87] (T2) | 1.16ms[45] (T2*) | 0.55ms[45] (T2*) |
| | Knee cartilage | 42.1ms[88] (T2) | 36.38ms[89] (T2) | 13.2ms[90] (T2) |
| | Achilles tendon | 2.18ms[91] (T2*) | 1.2ms[92] (T2*) | 8.5ms[92] (T2*) |
| | Trabecular bone | 4.5ms[93] (T2*) | 2.4ms[93] (T2*) | 1.2ms[93] (T2*) |
| | Ligament ACL | 11.9ms[94] (T2*) | 4.5ms[95] (T2) | |
| | MS lesion | 135ms[96] (T2) | 63.9ms[97] (T2) | 40ms[98] (T2*) |
| | Liver | 26ms[99] (T2*) | 20ms[100] (T2*) | |
| **SOFT TISSUES** | Subcutaneous Fat | 165ms[88] (T2) | 105.27ms[89] (T2) | 46.1ms[101] (T2) |
| | Muscle | 35.3ms[88] (T2) | 34.08ms[89] (T2) | 23ms[101] (T2) |
| | Heart | 52ms[102] (T2) | 46ms[102] (T2) | 17.4ms[103] (T2*) |
| | Gray matter | 84ms[104] (T2*) | 80ms[105] (T2) | 33.2ms[104] (T2*) |
| | White matter | 66.2ms[104] (T2*) | 56.4ms[106] (T2) | 26.8ms[104] (T2*) |
| | Kidney cortex | 112ms[107] (T2) | 121ms[108] (T2) | 108ms[108] (T2) |
| | Kidney medulla | 142ms[107] (T2) | 138ms[108] (T2) | 126ms[108] (T2) |
| **FLUID TISSUES** | CSF | 2500ms[109] (T2) | 2000ms[110] (T2) | 900ms[110] (T2) |
| | Blood arterial | 254ms[111] (T2) | 93ms[112] (T2) | 48.5ms(11.7T) (T2)[114] |
| | Blood venous | 181ms[111] (T2) | 60ms[113] (T2) | 20.1ms(11.7T) (T2)[114] |
| | Synovial fluid | 1210ms[88] (T2) | 767ms[88] (T2) | 324ms[101] (T2) |
| **WATER** | 3-4 seconds | | | |

## Background

To understand the physics behind an observed MRI signal, the following equation can be studied:

$$So = K\rho \sum crf(T1r, T2r),$$

Where $\rho$ is the total proton density and $c_r$, $T_{1r}$ and $T_{2r}$ are fraction of PD and effective time relaxation constants respectively. For a single component in an MRI image, the appropriate equation is:

$$So = KF\rho exp(-TE/T2)\ sin\alpha(1\text{-}exp(-TR/T1))/\ (1\text{-}cos\beta exp(-TR/T1))$$

Where TR is the sequence repetition time and TE is the time taken between central point of excitation process w.r.t rotation in XY plane and time at which the k-space is sampled at the same space. Here, $\alpha$ is the flip angle, which plays an important part in image acquisition[115]

Ultrashort Echo Time (UTE) and Zero Echo Time (ZTE) are advanced magnetic resonance imaging (MRI) techniques designed to capture signals from tissues with very short T2 relaxation times, with TE lower than 0.1 ms[116]. The principles behind these techniques can be understood through the following scientific explanations and equations:

1. **Increased Signal-to-Noise Ratio (SNR):**

   At higher magnetic field strengths (B0), the magnetization (M0) of nuclei in a tissue sample increases linearly with B0, as described by the equation:

   $M_0 \propto B_0$

   This increased magnetization results in a higher signal-to-noise ratio (SNR) in the acquired MR images. SNR is directly proportional to the square root of the number of spins contributing to the signal, which is itself proportional to the magnetic field strength. Therefore, at higher field strengths, UTE and ZTE sequences benefit from enhanced SNR, allowing for better image quality and shorter scan times.

2. **T1 and T2 Relaxation Times at High Field Strengths:**

   The T1 and T2 relaxation times of tissues can also change with magnetic field strength. The relationship between relaxation times and field strength can be described by the equations:

   $T_1 \propto 1/B_0$
   $T_2 \propto 1/B_0^2$

   These equations suggest that at higher field strengths, T1 values increase while T2 values decrease. For UTE and ZTE imaging, this has implications for tissue contrast and signal behavior. Tissues with very short T2 relaxation times, such as bone, tend to have even shorter T2 values at higher field strengths, making UTE and ZTE sequences even more crucial for their imaging.

3. **Improved Resolution:**

   The spatial resolution of MRI images is directly related to the magnetic field strength. The spatial resolution can be expressed as:

$Resolution \propto 1/B_0$

Therefore, higher field strengths enable higher spatial resolution, which is advantageous for UTE and ZTE sequences, especially when imaging fine structures or small regions of interest.

4. **Relationship of Susceptibility Effects with UTE and ZTE Sequences:**

   At high field strengths, susceptibility effects become more prominent. These effects arise due to variations in magnetic susceptibility among different tissues, leading to local field distortions. UTE and ZTE sequences are particularly useful in mitigating susceptibility artifacts because they acquire data during very short or zero echo times, reducing the impact of signal dephasing caused by susceptibility variations.

## Mechanism

The mechanism of UTE pulse sequences involves fast data acquisition, compared to traditional RF pulse sequences. The implementation of Ultrashort Echo Time (UTE) and Zero Echo Time (ZTE) sequences in magnetic resonance imaging (MRI) involves specific hardware considerations and sequence design principles to efficiently capture signals from tissues with very short T2* relaxation times. These sequences are characterized by the early activation of gradient fields at the beginning of signal acquisition, which results in a center-out trajectory in k-space. This trajectory is essential for capturing signals from tissues with rapid signal decay. To achieve this, short RF pulses with high peak power are employed to achieve the desired flip angle for excitation. Rapid switching between transmit (for RF excitation) and receive (for signal reception) modes is crucial for maintaining signal-to-noise ratio (SNR) during data acquisition. Diagram for UTE and ZTE sequences can be seen in Figure 1.

Optimal gradient performance is essential in UTE and ZTE sequences to detect short T2* species effectively. Gradients must ramp up quickly and maintain high amplitudes to encode spatial information efficiently. A high slew rate, which denotes the rate of change of gradient amplitude, is desirable for rapid spatial encoding. Data are collected and encoded using the slice selection gradient, first in one direction and then reversed, eliminating the need for pulse rephasing. Data acquisition continues while the gradient is ramped up, and a rapid gradient ramp-down is essential to prevent dephasing effects during signal acquisition. The k-space trajectory is often radial, although other trajectories like spiral, twisted, and cone may be used when suitable, depending on the imaging application.

UTE sequences can be acquired in both 2D and 3D modes, with half-pulse excitation being a common approach for 2D imaging[117]. These sequences efficiently capture images of tissues with short T2 and T2* relaxation times, making them valuable for various medical applications. Zero Echo Time (ZTE) sequences acquire data before the completion of RF pulses, with the gradient field already active, resulting in the center of k-space being crossed at zero echo time[118]. Reconstruction is necessary to fill in the center of k-space, as it is not directly sampled during the transition from transmission to reception mode. These techniques have valuable applications in musculoskeletal, neurological, and pulmonary imaging, where conventional MRI methods may be less effective. ZTE sequences, in particular, are known for providing a "silent MRI" experience with modifications, making them desirable in clinical settings. Hardware considerations play a vital role in the successful implementation of these sequences for achieving high-quality MRI images, involving the optimization of gradient performance, RF pulse design, and sequence parameters.

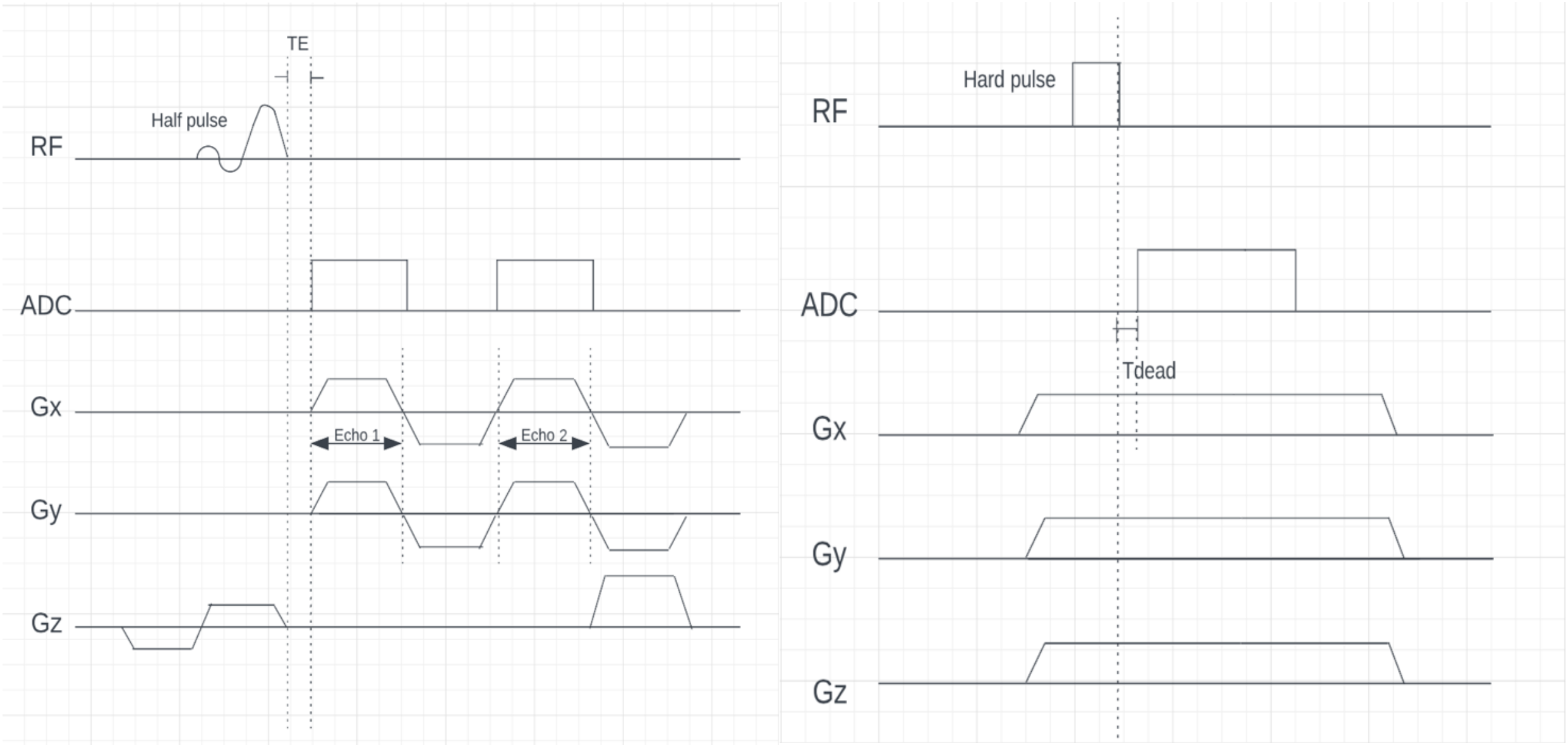

Figure 1: UTE and ZTE pulses respectively. Note that data acquisition starts before the completion of the RF pulse.

## Practical Implementation of UTE and ZTE Techniques

### Musculoskeletal System (MSK) Imaging

Advanced MRI techniques in conjunction with ultra-high field strengths offer exceptional capabilities for rapid, high-resolution imaging of the musculoskeletal system, including tendons and bones[119]. The human bone, a highly complex structure, primarily comprises organic matrices fused with hydroxyapatite calcium phosphate (HA) crystals, divided into two major tissue types: cortical bone and trabecular bone. Non-UTE techniques like Water and Fat Suppressed Projection MRI (WASPI) have been utilized to suppress fluid signals and emphasize solid matrix signals from bones[120]. Cortical bone, constituting 80% of the skeletal weight, plays a crucial role in diagnosing osteoporosis due to cortical shell thinning, pore expansion, and mineral-matrix depletion[121].

Custom techniques like Inversion Recovery UTE (IR-UTE) sequences help in suppressing long T2 signals from soft structures, enhancing image focus. The Dual Inversion UTE (DIR-UTE) technique[122] has been employed successfully for cortical bone detection in cadaveric and in-vivo studies at 3T, achieving excellent signal suppression and contrast-to-noise ratios for Achilles tendon and cortical bone detection. SWIFT (Sweep Imaging with Fourier Transformation)[123, 124] is an emerging MRI technique known for its short acquisition times, sensitivity to short T2 signals, reduced signal dynamic range, near-zero echo time (TE), and minimal acoustic noise. SWIFT imaging has proven its effectiveness in visualizing cortical bone compared to standard Gradient Echo Sequences (GRE) in ex-vivo studies, highlighting its advantages in terms of clarity.

In UTE and ZTE pulse sequences, radial trajectory is commonly employed. In-vivo experiments at different field strengths of 3T and 7T have been conducted to investigate the impact of radial UTE[92, 125]. Herrmann et. al.[125] utilized various TE times and achieved faster acquisition times compared to dual-echo Cartesian sequences, demonstrating its efficiency. Han et. al.[92] demonstrated the feasibility of 3D radial UTE at 7T for Achilles tendon imaging, showcasing the visualization of minute internal fiber structures.

Employing spiral trajectories has also improved signal detection, countering radial trajectory limitations. Spiral UTE has efficiently detected short T2 components in white matter, deep layers of cartilage, and cortical bone, showing promise[126-128]. A combination of dual-band UTE and dual-echo UTE has yielded improved SNR and CNR[129]. A comparison between ZTE and UTE at 7T showed similar contrast properties for both techniques, with shader artifacts detected in the case of ZTE[86]. High-resolution imaging of the head, ankle, and knee with UTE sequences demonstrated efficient visualization of fascicular structures[86].

While UTE sequences exhibited excellent results with a minimal flip angle, ZTE sequences with zero echo time required additional suppression of padding materials to mitigate signal artifacts. 3D ZTE imaging at 7T yielded high SNR results and spatial resolution for various musculoskeletal regions[130]. These images effectively captured short T2 signals from bones, cartilages, tendons, and ligaments, providing clarity and distinction at bone-tissue interfaces. The rapid switching of gradients contributed to artifact-free imaging, offering a comprehensive view of the musculoskeletal system.

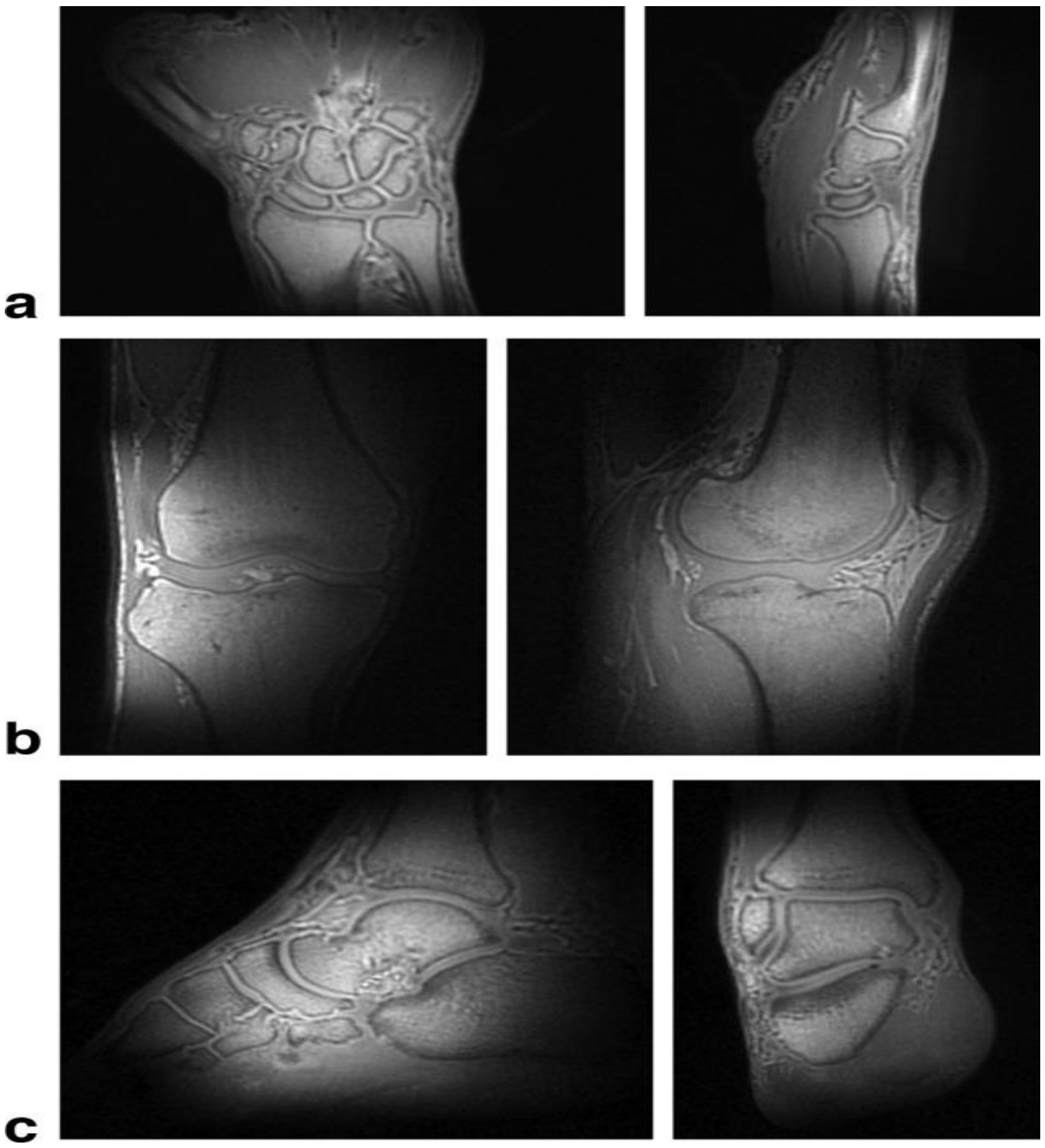

Figure 2: ZTE images of joints of (a) wrist (b) knee (c) ankle. Delineation of joints against blood vessels and other layers were clearly depicted with high SNR. No artifacts associated with fast RF switching by Weiger[130] with permission

In-vivo Zero Echo Time (ZTE) imaging of the glenohumeral joint was conducted by Breighner et. al.[131] at both 1.5T and 3T magnetic field strengths. The field of view (FoV) ranged from 20 to 30 cm, employing a minimal flip angle of 1 degree, with scan times varying from 4 to 6 minutes. This ZTE imaging approach proved effective in detecting subchondral bone and bone resorption around the MR hardware. Schieban et. al. also conducted a ZTE imaging study, which consistently yielded valuable results[132].

Chang et. al.[133] performed a comparative analysis between Fast Spin Echo (FSE) and Ultrashort Echo Time (UTE) techniques for imaging the Achilles tendon. Figure 3 illustrates this comparison, featuring a proton density FSE image with a TE of 6.6 ms, displaying no discernible signal from the Achilles tendon. In contrast, the 3D UTE cones imaging utilizing a TE as short as 30 µs clearly depicted the signal originating from the internal structure of the repaired Achilles tendon. The third image, a difference map between the FSE and UTE images, vividly illustrates the missed signals in the FSE image. This comparative study underscores the superior capability of UTE imaging in capturing rapid tissue signals, particularly in challenging anatomical regions like the Achilles tendon.

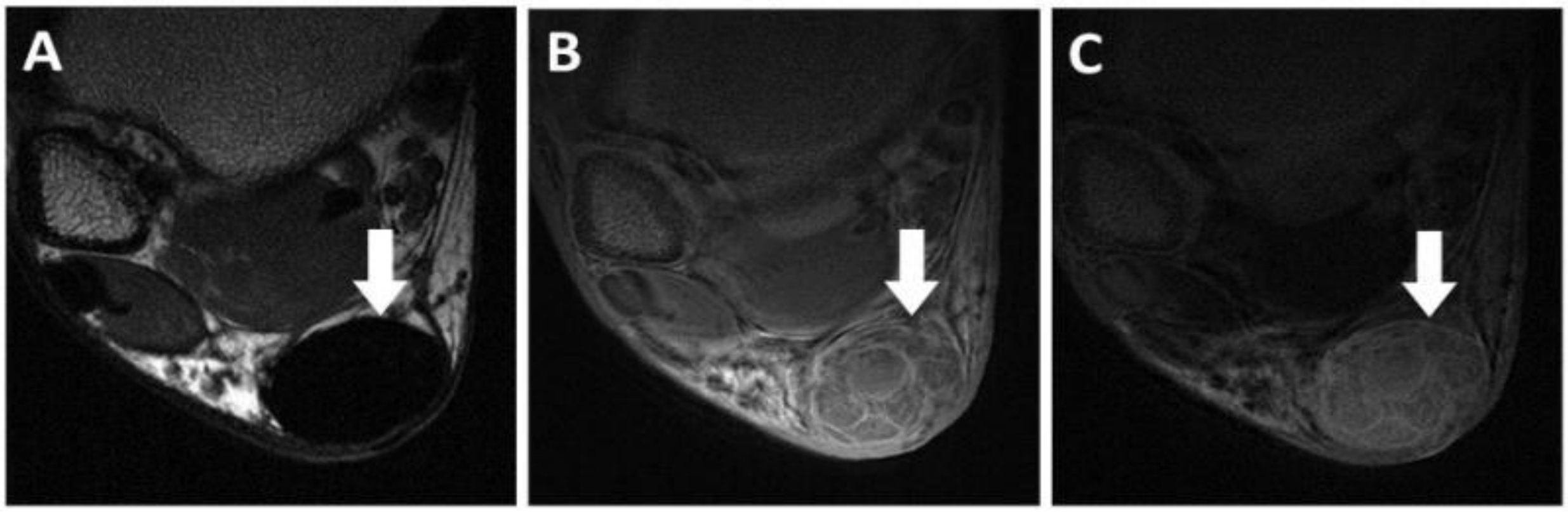

Figure 3: Axial images of Achilles tendon repair. (a): PD FSE image with TE = 6.6 ms. No signal is depicted within the Achilles tendon. (b) 3D-UTE-Cones image using TE = 30 us. Void signal from Achilles tendon is finally replaced by successful visualization of the internal structure. (c): Subtraction of first and second image by Chang[133] with permission.

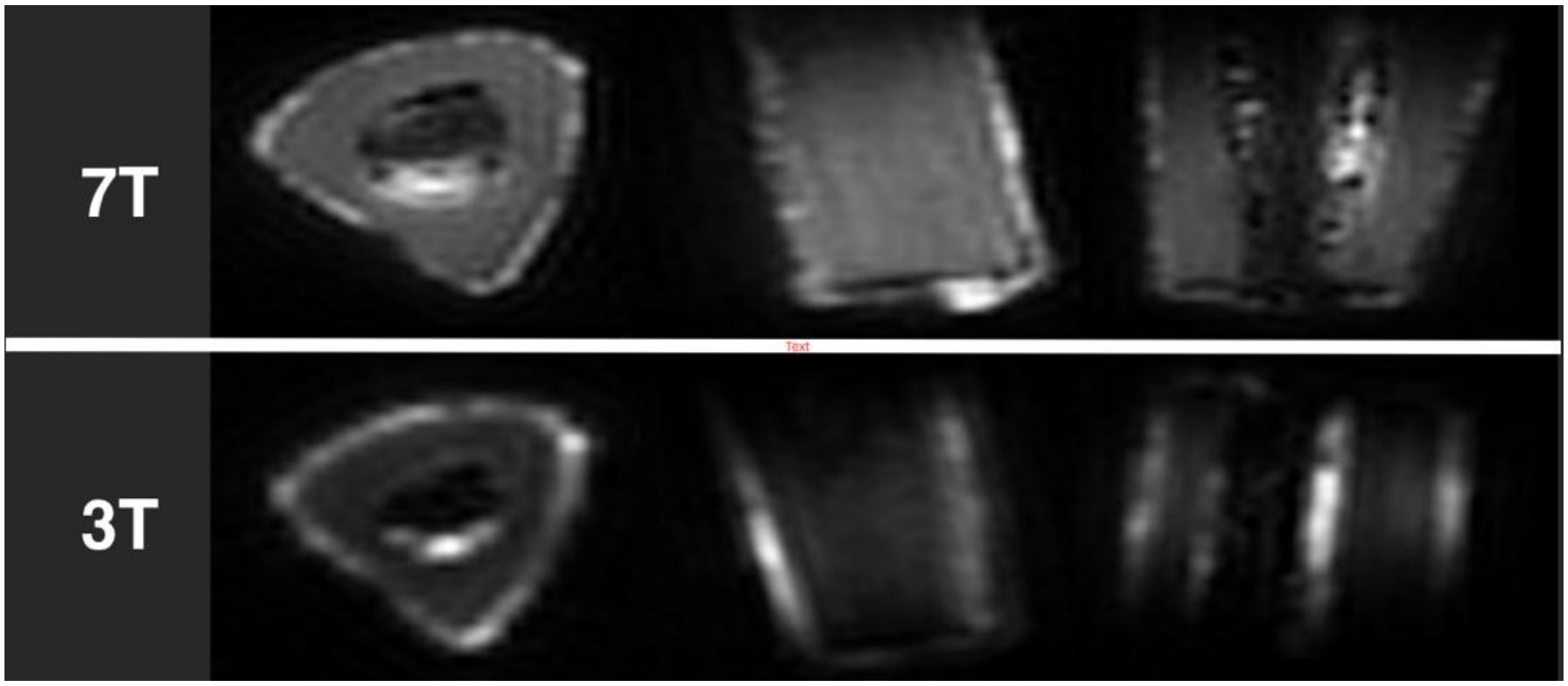

Figure 4: Imaging of cadaveric radius at 7T (top row) and 3T (bottom row) using an in-house saddle coil. Signal is detected from the cortical bone with 7T giving a better SNR. Scan time was found to be 4:40 min by Krug et. al.[45] with permission.

Cortical bone, owing to its inherently short T2 relaxation time, often presents as a void signal when imaged using conventional MRI techniques. Consequently, the utilization of Ultrashort Echo Time (UTE) and Zero Echo Time (ZTE) imaging becomes imperative for effectively visualizing cortical bone. Krug et. al.[45] conducted a study involving the imaging of mid-diaphyseal sections from five fresh cadaveric specimens of the distal tibia, employing different field strengths of 3T and 7T. A saddle-coil configuration was employed for imaging, as depicted in Figure 4. The scan duration was determined to be 4 minutes and 40 seconds, with an isotropic voxel size of 1mm and an isotropic resolution of 0.8mm. This study underscores the significance of utilizing UTE and ZTE sequences for enhanced cortical bone imaging, particularly at higher field strengths.

The correlation between Signal-to-Noise Ratio (SNR) and magnetic field strength was substantiated, with 7T imaging demonstrating a 1.7 times higher SNR compared to 3T. Additionally, it was emphasized that T2* relaxation time exhibits an inverse relationship with field strength, as 7T imaging yielded smaller T2* values in comparison to 3T. This phenomenon primarily arises due to pronounced magnetic field inhomogeneities and susceptibility effects associated with higher field strengths.

Emerging techniques like UTE-MT (Magnetization Transfer), as described by Yang et. al.[134], have proven valuable in diagnosing early cartilage degeneration. The study was conducted at 3T, featuring an extremely short TE of 0.032 ms, a Field of View (FoV) measuring 8 cm, and a total scan time of 1 minute and 48 seconds. UTE-MT was compared with UTE-T2* mapping and conventional T2 mapping. The diagnostic efficacy of UTE-MT surpassed that of UTE-T2* mapping and conventional T2 mapping, a finding further corroborated by its correlation with Mankin scores. UTE-MT emerges as a highly valuable technique for assessing cartilage health, offering significant potential in early degeneration detection.

## Lung Imaging

Traditionally, MRI was not the preferred imaging modality for lung evaluation, mainly due to challenges related to low proton density, motion artifacts, and longer acquisition times compared to CT scans. However, recent advancements aimed at mitigating these challenges, combined with MRI's inherent advantages of providing superior visual and quantitative assessments without ionizing radiation, have sparked renewed interest in MRI for lung imaging. One lung condition that has garnered attention for MRI applications is Cystic Fibrosis, a genetic disease primarily affecting the lungs[135]. MRI offers the unique capability to distinguish between scarred tissue and active inflammation, making it particularly valuable in diagnosing Cystic Fibrosis. Additionally, MRI can detect lung edema and emphysema[136]. Short echo time techniques are advantageous for maintaining image quality in lung imaging. As a result, assessments for various lung ailments, including non-small cell lung cancer, diffuse lung disease, and pulmonary edema, have become feasible[137].

However, implementing basic 3D radial Ultrashort Echo Time (UTE) sequences presents challenges due to lower Signal-to-Noise Ratio (SNR) efficiency compared to their 2D counterparts. To address these issues, techniques such as variable density readout gradients, respiratory gating, and limiting the field of view can be employed to enhance image quality. The application of these techniques has led to improved SNR, enabling the visualization of lung parenchyma, airways, and fissures. Different gas mixtures can also be introduced simultaneously to enhance ventilation images in cases of obstructed airflow[138]. The Zero Echo Time (ZTE) technique can effectively replace the use of fluorinated gases[139].

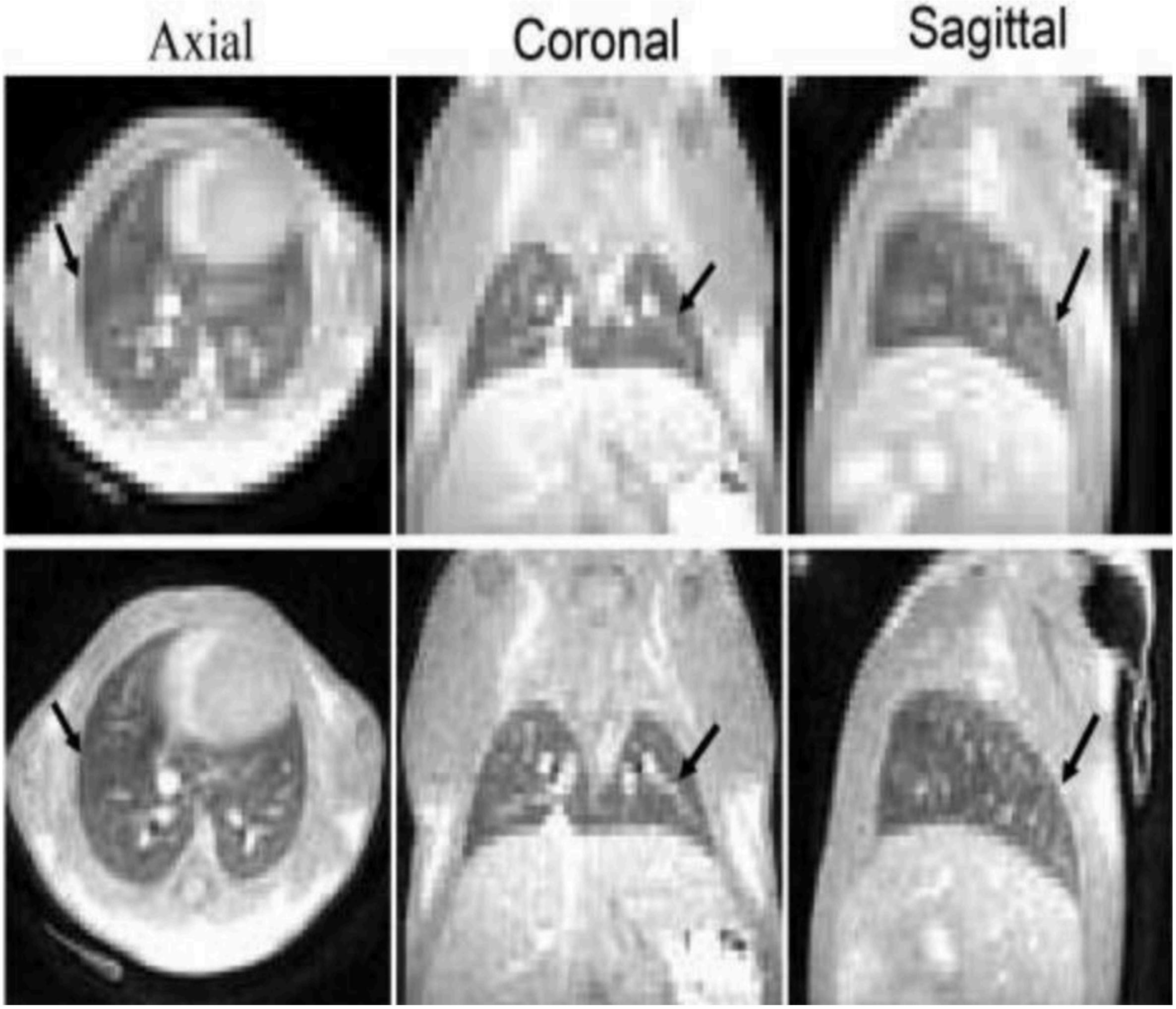

Figure 5 depicts axial, coronal, and sagittal images acquired using spherical and ellipsoidal k-space coverage. The ellipsoidal UTE image exhibited higher resolution, as indicated by black arrows, enabling better imaging of lung boundaries and blood vessels[140]

In an in-vivo study conducted by Guo et. al.[140], T2* and T1 measurements of lung parenchyma were performed at 7T using a modified UTE sequence. The study involved imaging 12 mice using an optimized UTE sequence with ellipsoidal k-space coverage to reduce scan time while maintaining high-resolution imaging. This approach resulted in higher resolution images, facilitating accurate measurement of T2* relaxation times and improved delineation of blood vessels from lung parenchyma. Images of the murine pulmonary region in axial, coronal, and sagittal planes showcased the enhanced resolution achieved with ellipsoidal UTE acquisition. The second image demonstrated higher SNR, depicting clearer boundaries of the lung and blood vessels without a significant increase in scan time.

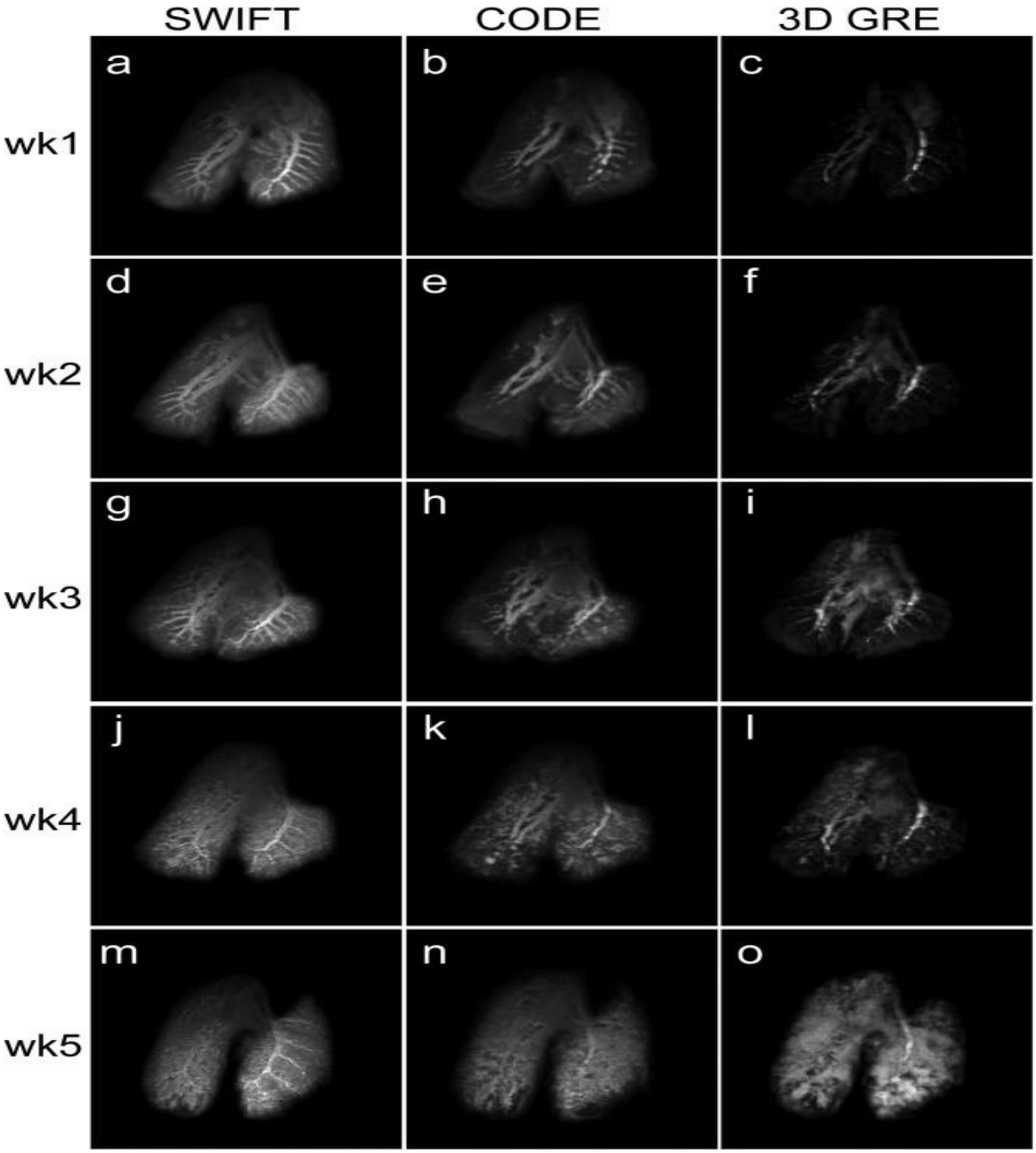

Figure 5: MIP images of metastasis of lung due to breast cancer cells using SWIFT, CODE and GRE techniques. Images from SWIFT clearly detected minute structures around blood vessels better than other techniques. From Kobayashi et. al.[141] with permission.

Kobayashi et. al. implemented the Sweep Imaging with Fourier Transformation (SWIFT) technique for evaluating breast cancer metastasis to murine lungs. The study utilized a 9.4T animal MR scanner with specific gradients and a quadrature coil for transmitting and receiving. SWIFT imaging parameters included a TE of 3 µs, a field of view (FOV) of 40 × 40 × 40 mm3, a nominal flip angle of 10 degrees, and a scan time of 60-80 minutes. SWIFT imaging was compared with CODE (concurrent dephasing and excitation) and conventional Gradient Echo Sequences (GRE). The study used MIP (maximum intensity projection) images to monitor metastasis growth and visualize vasculature. SWIFT's short TE and resistance to susceptibility artifacts made it more effective in detecting small structural changes compared to the other methods.

## Dental Imaging

UTE and ZTE sequences have emerged as excellent alternatives to traditional CT and other imaging methods for dental imaging. They offer the capability to detect both soft tissues and hard components simultaneously, providing clear differentiation between various dental structures such as enamel, dentin, cementum, and pulpa[142]. These techniques significantly surpass conventional MRI and CBCT methods. Leveraging higher field strengths, such as 7T, can further enhance image contrast and resolution.

An ex-vivo 3D ZTE MRI of a horse's tooth at 9.4T demonstrated the potential of these techniques[142]. The ZTE image (a) of the horse tooth closely corresponds to the actual photograph of the top surface (b), showcasing excellent visualization of the solid components of the extracted tooth. Additionally, a compromised tooth is depicted in (c), highlighting the damage evident in the compromised pulpa at the mid-section compared to the healthy pulpa in the lower section, as indicated by arrows. The isotropic resolution was 235 $\mu m^3$ for (a) and 470 $\mu m^3$ for (c), with acquisition times of 13 minutes and 90 minutes, respectively.

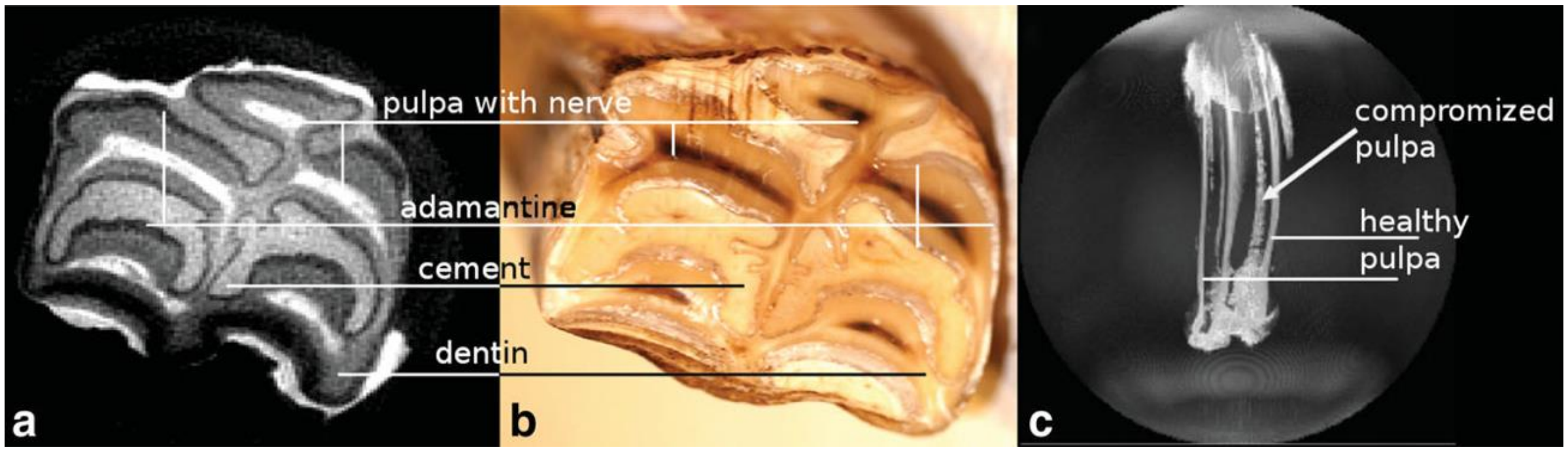

Figure 6: Imaging of Horse Tooth. 3D ZTE MRI of a healthy (a,b) and compromised (c) tooth was taken. Good contrast between the layers was discovered with clear distinction between healthy and compromised pulpa by Hovener[142] with permission.

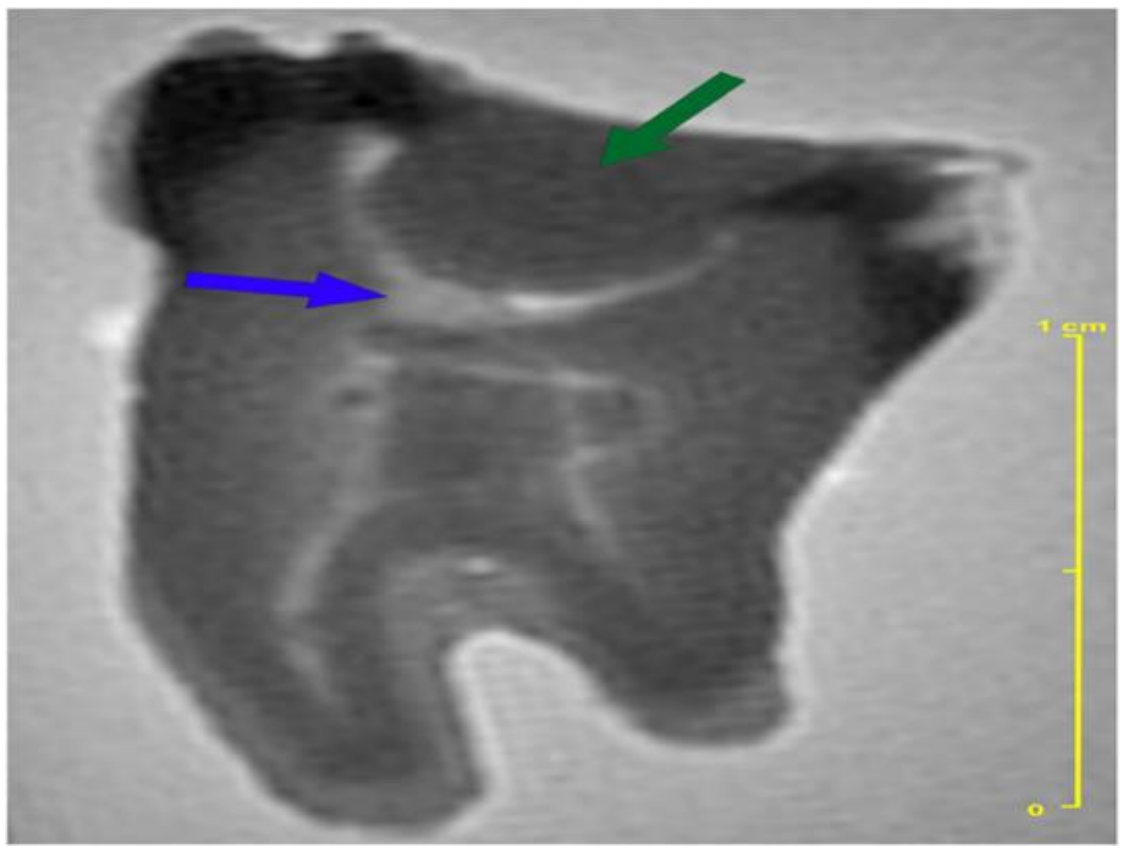

Figure 7: ZTE image of Wisdom Tooth is seen above. Restoration material is detected by green arrow, while blue arrow shows detection of CPC. Screening of biomaterials with ZTE is found to be a reliable tool for biomaterial detection by Mastrogiacomo[143] with permission.

Figure 8 displays an in-vivo ZTE image of a human tooth captured at 11.7T during a study conducted by Mastrogiacomo et. al.[143]. The tooth had undergone pulp capping using Calcium Phosphate-Based Compositions (CPC) and had glass ionomer restoration material applied. ZTE imaging successfully detected the short T2 relaxation of CPC, which typically goes undetected with conventional MR sequences (blue arrow in Figure 8). Additionally, the glass ionomer restoration material was clearly visualized (green arrow). ZTE proves to be a reliable tool for imaging teeth and dental materials, enabling the screening of biomaterials and the assessment of tooth integrity through MRI. The matrix size parameters were set at 256 × 256 × 256, with a flip angle of 3 degrees, and the acquisition time totaled 13 minutes and 46 seconds.

The SWIFT technique also demonstrated promising results in imaging fast relaxing and short T2* components of teeth[144]. Unlike conventional UTE sequences, SWIFT utilizes a swept RF excitation while concurrently acquiring signals in the presence of field gradients. SWIFT offers the advantage of imaging truly ultrashort T2 components with relatively low peak RF amplitude, reducing the strain on MR hardware. This technique is a preferred alternative for imaging calcified dental tissues and can be used for comparisons with traditional techniques like X-ray modalities and histological sections. In-vitro MRI experiments were performed at a 9.4T Magnex Scientific scanner with a flip angle of 15 degrees, acquiring data in 256 pulse gaps with a gap duration of 6 ms each. High-resolution images were obtained with a scan time of 24 minutes. A comparison of different dental imaging methods conducted by Idiyatullin and Garwood et. al. revealed that SWIFT and CBCT yielded the best results. The appearance of composite resin restoration materials varied in brightness or darkness due to different mineral concentrations, posing diagnostic challenges. SWIFT and CBCT provided better visualization of dentin compared to GRE imaging, where all hard tissues exhibited no signals. Notably, SWIFT had an advantage over CBCT as it enabled the visualization of accessory canals within the apical one-third of the root apex[144].

## Neural System Imaging

Myelin, a vital component of the neural system, comprises approximately 14% of the wet mass of white matter and plays a critical role in neural transport, conduction speed, and reduced axonal energy requirements[145]. Myelin deficiency is a primary cause of neural diseases such as Multiple Sclerosis (MS) and schizophrenia, resulting in demyelination and irreversible symptoms, including speech decline, balance issues, and cognitive impairment[146]. Consequently, assessing myelin is crucial for detecting central nervous system (CNS) abnormalities. Traditional methods to detect myelin, such as x-ray diffraction, non-linear optical techniques, and optical microscopy, are destructive and limited to animal studies. MRI, being non-destructive, has become the leading system for imaging these tissues, with ZTE imaging being increasingly employed to evaluate skull lesions resulting from past head trauma[147].

Myelin imaging is of great interest for both clinical and research purposes, especially in the context of neurological disorders[148]. Given that myelin exhibits transverse relaxation properties similar to those discussed for the musculoskeletal system, the use of ultra-short echo time (UTE) sequences is essential. Gray matter lesions are a common feature of MS and significantly contribute to the progression of the disease[149]. Conventional MRI sequences often lack the specificity required to assess pathological substrates in MS and fail to provide accurate details about brain damage. These sequences are susceptible to image artifacts and interference from fatty tissue[150]. UTE sequences excel in distinguishing demyelination and remyelination, crucial for disability diagnosis[151]. The adiabatic inversion recovery UTE technique can effectively detect myelin in the white matter of the brain[152]. UTE-MT (Magnetization Transfer) is a valuable tool for detecting alterations in cortical areas[153]. Studies conducted by Ma et. al. at both 3T[152] and 7T[154] utilized IR-UTE techniques for in-vivo brain imaging. The 3T study revealed areas of abnormality that appeared normal in conventional FLAIR and T2*-weighted sequences, critical for accurate Multiple Sclerosis diagnosis. The 7T study, utilizing the cuprizone (CPZ) model to detect demyelination, featured various echo times (TEs) and provided excellent contrast, suppressing long T2 signals from white matter and gray matter. Signals from the skull, not detectable in conventional FSE sequences, were captured in the IR-UTE sequence.

Larson et. al. compared ZTE and UTE images for brain imaging and found that both sequences provided similar contrasts between gray and white matter, with the added advantage of detecting signals from the cortical bone of the skull[86]. However, ZTE images exhibited more prominent artifacts due to padding material. Weiger et. al. demonstrated the results of 3D ZTE imaging, using birdcage coils, on healthy volunteers in the head area, showcasing B1 non-uniformity in the head leading to smooth intensity variations related to the usage of birdcage coils[130].

The impact of field strength on image quality is illustrated in Figure 10, showing the difference between in vivo head images taken at 1.5T and 7T. The 7T image provided superior visualization of a microbleed in the brain due to trauma, which was less distinct at 1.5T[155]. This example underscores the importance of early diagnosis and precise imaging for medical procedures.

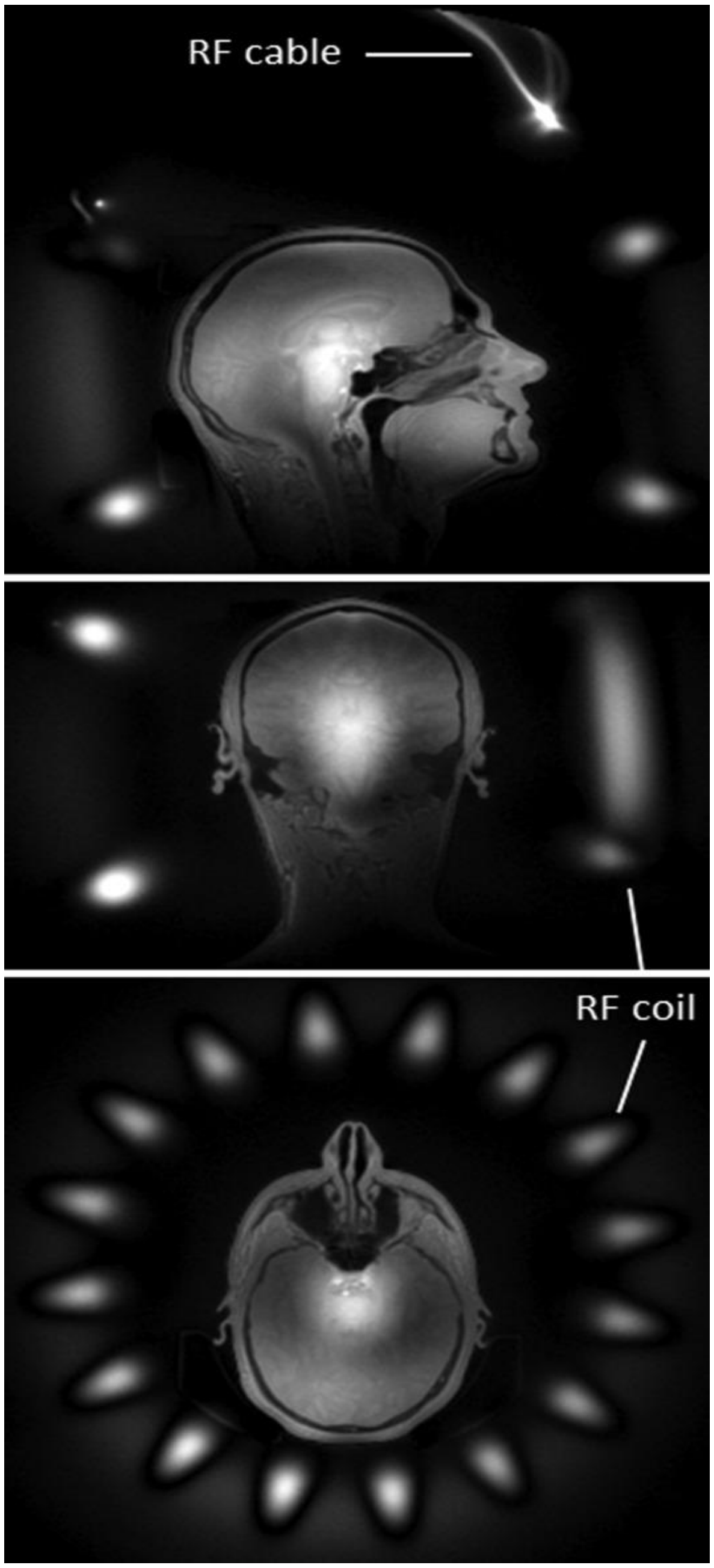

Figure 8: ZTE images of a human head using a birdcage coil. Three different orthogonal slices are shown. Non-uniformity associated with 7T spotted with image exhibiting proton density (PD) weighting by Weiger[130] with permission.

Figure 11 demonstrates the difference between UTE and various scans at 3T, highlighting UTE's ability to provide excellent anatomic correlation, particularly in spotting Schmorl's node in the superior endplate of L3[156]. UTE outperformed other scan types, including Sagittal T2-W FS,

T1W FSE, and 3D slab-selective UTE images, in detecting this feature. The total scan time for these scans was 9 minutes.

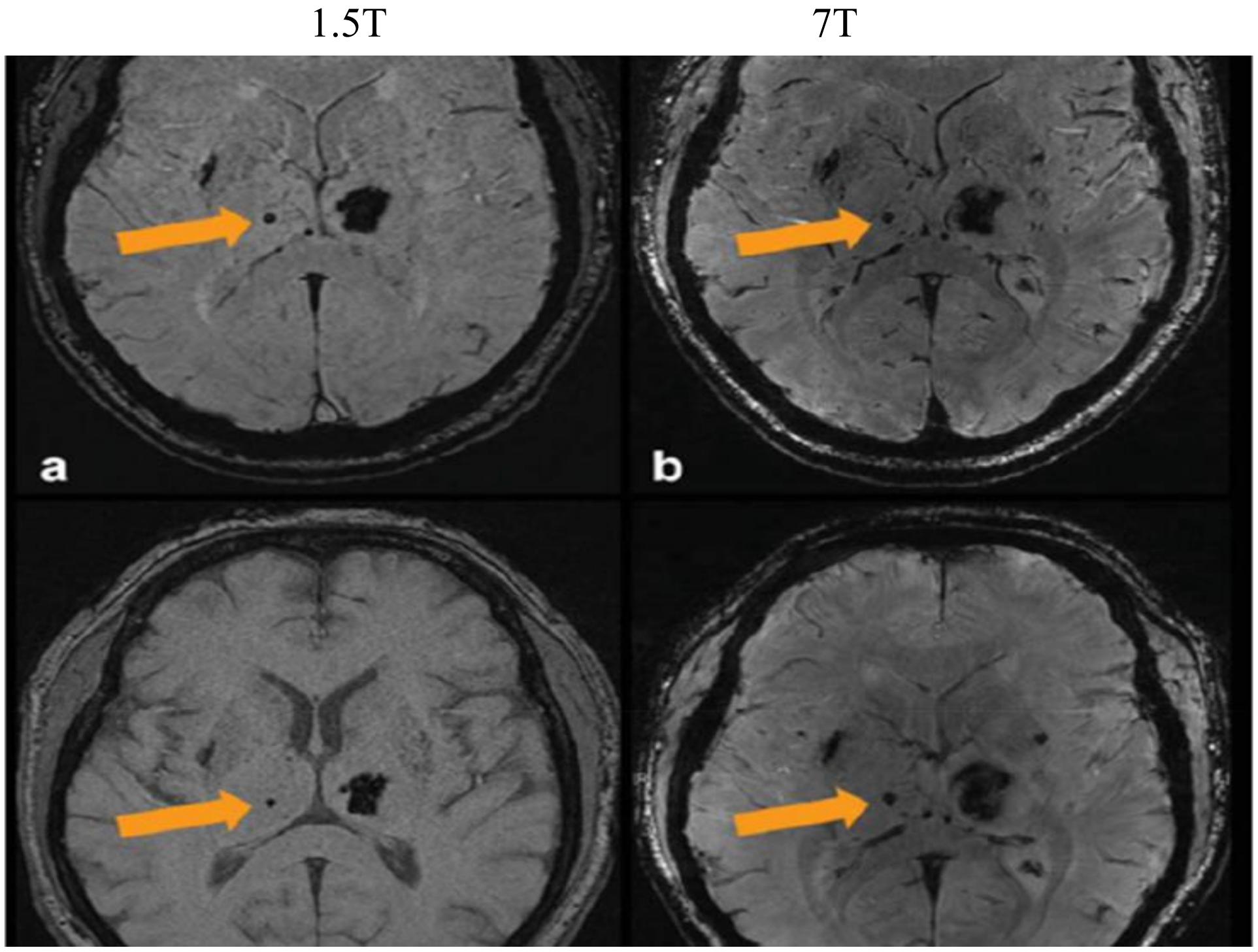

Figure 9: MRI of Microbleed at 1.5T and 7T. (a) is clinical SWI at 1.5T and (b) at 7T. Cerebral microbleeds are easier to detect at 7T with addition of surrounding detail shown by Theysohn[155] with permission.

Figure 11: Sagittal T2-W FS and T1-W FSE images (A,B) and sagittal (C), coronal (D) and axial (E,F) 3D slab selective UTE images from a 42-year-old male patient with low back pain (T-12-

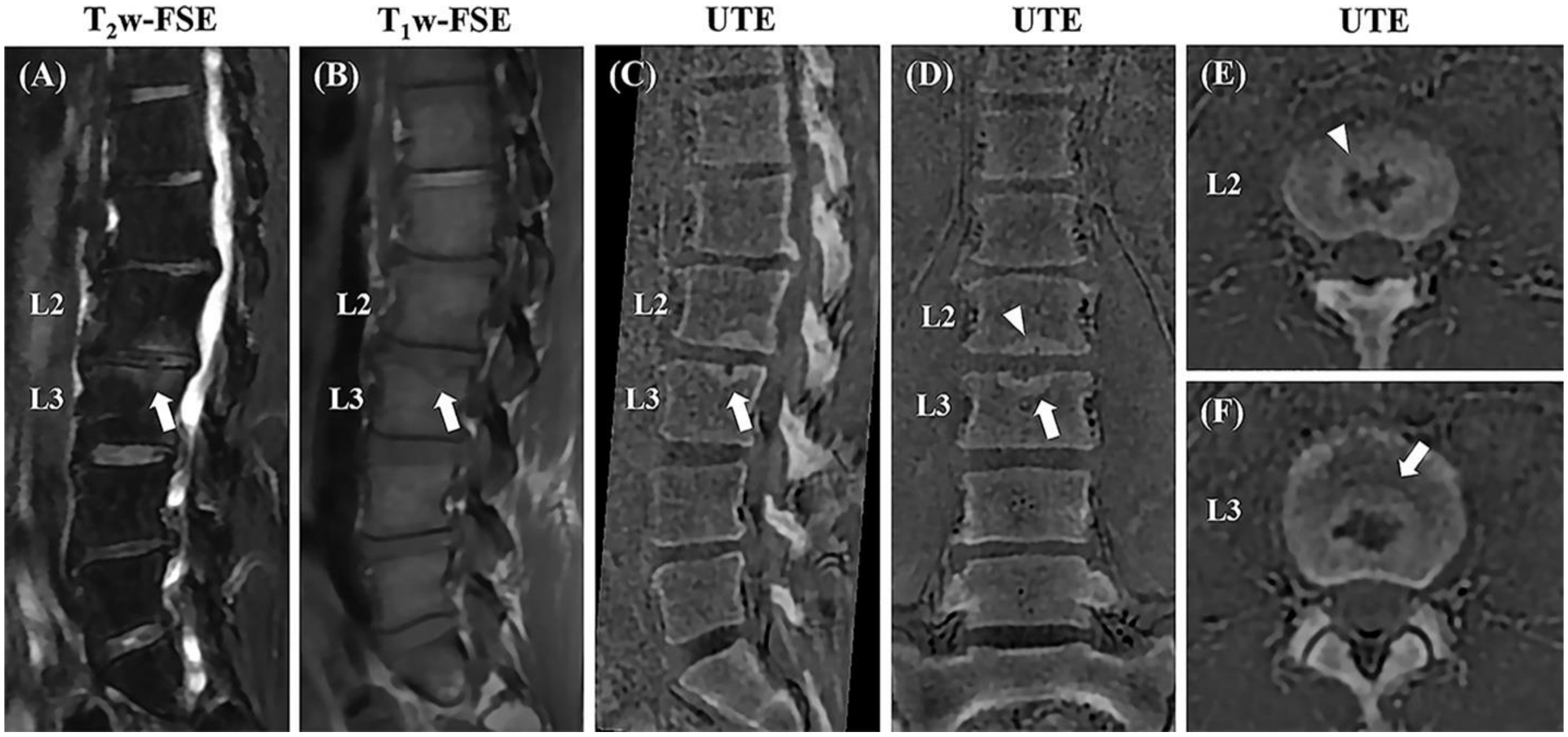

L5). Image demonstrates excellent anatomic correlation with 3D slab selective UTE sequence as shown in Afsahi[156] with permission.

Table 2: Describing different sequences discussed and its scan time and tissue characteristics

| **Sequence Type** | **TE time** | **Author** | **Avg. Scan Time Eff.** | **Contrast** | **Field Strength** | **Human/ Animal Studies** | **Tissue Characteristics** |
|---|---|---|---|---|---|---|---|
| **Radial UTE** | 8-12000 µs | Du et. al.[126] | Low 5.1-9.5 min | Low | 1.5T | Brain white matter, patellar, femoral, tibial cartilage, meniscus and cortical bone | Highest SNR for bone, with least for brain. Fastest scan time for cartilage |
| | 60-100 µs | Herrmann et. al.[125] | 6:20 min-1:14hr | Low | 3T | Head, tibia, lung | small anatomical structures in the nasal cavities are well delineated, accurate estimation of T2* of different tissues in tibia, blood vessels and blood paths in lungs delineated |
| | 229 µs | Han et. al.[92] | 18 min | Low | 7T | Achilles Tendon | high signal received from tendon, depicting fascicular pattern |
| | 64-2048 µs | Krug et. al.[45] | 14 min | | 7T | Human cadaveric tibiae | Clear demonstration of detection of MR signal from cortical bone |
| **IR-UTE** | 0.032-0.8 ms | Ma et. al.[152] | Low<br>7 min | High | 3T | Bovine myelin lipid powder, human brain, rat thoracic spinal cord | Significantly correlated positively with cortical bone stiffness, strength, and toughness to fracture |
| | 0.02-2ms | Ma et. al.[154] | 100 min | High | 7T | Mouse brain | Sharp contrast detected between myelin protons with appreciable low T2 suppression |
| **Dual IR UTE** | 8 µs | Du et. al.[122] | High<br>5 min | High | 3T | Patella and spine | Suppresses signals from the superficial layers of articular cartilage and fat |
| **Spiral UTE** | 8-12000 µs | Du et. al.[126] | High<br>3.8-7.5 min | High | 1.5T | Brain white matter, patellar, femoral, tibial cartilage, meniscus and | Highest SNR for bone, with least for brain. Fastest scan time for cartilage, higher SNR efficiency compared to Radial UTE |
| | 32 µs | Wan et. al.[127] | 4.65 min | High | 3T | Human knee | Deep cartilage and PCL detect with good contrast |
| | 10ms | Qian et. al.[128] | 6.4min (64 spirals) | Low | 7T | Brain | Delineating of blood vessels (bright) with less undersampling artifacts as compared to GRE image |
| **UTE-MT** | 0.032-14.7ms | Yang et. al.[134] | High | Low | 3T | Human anterolatera | Strong correlation with histological |
| | | | 5:13 min | | | l condyle specimens | grades of cartilage degeneration |
| **Anisotropic UTE** | 0.075-1.73ms | Guo et. al.[140] | | Low | 7T | Murine lungs | Sharper lung boundaries and clearer blood vessels detected |
| **3D Slab-Selective UTE** | N/A | Afsahi et. al.[156] | 9min | High | 3T | Human Spinal Cord | Better detection of Schmorl's node |
| **UTE with periodically applied fat suppression pulses** | 2.3ms | Larson et. al.[86] | 3:30-3:42 min | High | 7T | Human head | good contrast for suspicious demyelinated lesions |
| **ZTE** | N/A | Weiger et. al.[130] | 2–3 min | Low | 7T | Human head, joints | Smooth intensity variations observed in head. Joint images exhibit rich anatomical detail at high SNR |
| | N/A | Larson et. al.[86] | 5 min | | 7T | Human head | good contrast for suspicious demyelinated lesions |
| | N/A | Hovener et. al.[142] | 13/90 min | High | 9.4T | Horse tooth | Delineation of layers of teeth, defects readily depicted |
| | N/A | Mastrogiacomo[143] | 13:46 min | Low | 11.7T | Human tooth | Successful detection of CPC and Glass ionomer restoration |
| **ZTE with bias-correction algorithm** | 8 µs | Breighner et. al.[131] | High<br>4-6 min | Low | 3T | Human glenohumeral joint | Visualization of subchondral bone depression and bone resorption around hardware |
| **Enhanced flip angle ZTE** | N/A | Schieban et. al.[132] | High | Low | 4.7T | Pork's hock | Signal from bone observed, otherwise not spotted by GRE |
| **SWIFT** | N/A | Idiyatullin et. al.[123] | High | Low | 4.7T | Bovine tibia | Cortical bone observed in SWIFT image which was not detected using gradient echo imaging |
| | N/A | Idyatullin et. al.[144] | 24 min | Low | 9.4T | Human Teeth | SWIFT can visualize otherwise not observable anatomical details such as accessory canal within the apical one third of the root apex |
| | 3 µs | Kobayashi et. al.[141] | 60-80 min | | 9.4T | Murine lungs | Less susceptibility artifacts and detection of minute structural changes |
| **Multi-Band SWIFT** | N/A | Idiyatullin et. al.[124] | 4 min | Low | 4T | human mandible | discontinuity in the (hypointense) mandible detected, possibility of tumor |
| **WASPI** | N/A | Wu et. al.[120] | Low | High | 4.7T | Rats, chicken bone marrow | Artifacts were significantly corrected after image reconstruction |

## Conclusion

UTE and ZTE sequences have unequivocally demonstrated their utility in imaging solid or semi-solid tissues characterized by short echo times due to strong dipole-dipole interactions. This versatility extends across diverse areas of interest, ranging from the head to the lungs and beyond. These sequences excel in imaging short T2* components, including cortical bone, tibia, and achieving near-silent acquisitions through minimal gradient switching in ZTE, thereby addressing long-standing issues in MRI, such as void signals from short T2* species and the high decibel levels associated with operation.

Techniques like SWIFT imaging, IR-UTE, and the application of various trajectory types have notably enhanced results by improving image focus and suppressing long T2* structures, ultimately leading to superior signal detection. The use of ellipsoidal UTE sequences has facilitated the improved detection of blood vessels and other components without an additional increase in scan time. The combination of dual band UTE and dual-echo UTE has yielded enhanced signal-to-noise ratio (SNR) and contrast-to-noise ratio (CNR). There is also a confirmed correlation between SNR and field strength, with a 1.7-fold increase at 7T compared to 3T.

Dental MRI is experiencing increased adoption due to results comparable to the cone beam computed tomography (CBCT) method. Furthermore, ultra-short echo time sequences aid in the detection of myelin deficiency, a primary factor in Multiple Sclerosis. However, leveraging the advantages of ultra-high field strengths and ultra-short echo times does present certain disadvantages and technical challenges. There remains a limited availability of efficient radiofrequency coils designed to detect short-T2 species in conjunction with ultra-high field strengths. Shader artifacts associated with ZTE sequences can adversely affect white matter contrast, and physiological issues like dizziness and nausea may occur. Breathing artifacts pose concerns in lung imaging.

Nonetheless, ultra-high field strengths, particularly at 7T or higher, have demonstrated immense potential for localized scanning in areas such as the head, knee, ankle, teeth, tibia, and spinal cord, rapidly emerging as a promising tool for lesion detection and the imaging of other short T2* species. This potential can be further harnessed through the application of specialized trajectories like spiral or ellipsoidal acquisitions, enabling efficient k-space filling. With ongoing advancements and future progress, ZTE/UTE sequences at high and ultra-high magnetic field strengths are poised to significantly expand the domain of MRI applications, including full-body scans, while adhering to local Specific Absorption Rate (SAR) considerations.

## Discussion

Recent advancements in the field of MRI have introduced several innovative techniques, including Ultra-short Echo Time (UTE), Sweep Imaging and Fourier Transformation (SWIFT), Water and Fat Suppressed Projection MRI (WASPI), which are particularly effective in studying short T2 signals like 1H signals emanating from areas such as cartilage and bone[157]. However, heightened sensitivity to these signals also amplifies the impact of background 1H signals originating from plastics and padding materials, significantly affecting image quality. These undesired fast-relaxing background signals can emerge both from external and internal sources within the RF coil. Consequently, specific hardware considerations are necessary for optimizing efficiency. One common approach to address this challenge is to employ pulse sequence techniques and specialized selection methods related to gradients and Radio Frequency. Nevertheless, due to inherent limitations, minimizing and shielding all potential sources of background signals becomes imperative.

Table 2 provides an overview of various techniques discussed in this paper and illustrates how different parameters influence scan time and tissue characteristics. As previously mentioned, increasing the magnetic field strength (B0) enhances the signal-to-noise ratio (SNR) but also results in increased B1 inhomogeneity, potentially violating Specific Absorption Rate (SAR) constraints due to elevated power deposition. Consequently, the necessity for multi-channel RF transceiver systems for B1 shimming and SAR management arises, adding complexity. Thus, the choice of the optimal field strength becomes highly dependent on the specific application. To attain superior image quality, it is essential to maximize slew rates and gradient performance. An ideal RF receiver system should exhibit rapid recovery capabilities. Particularly in sequences with reduced echo times like ZTE, the rapid switch between transmit and receive modes gain paramount importance. Hence, higher peak RF transmitters are desirable for this purpose. Substantial oversampling is preferred to achieve a significantly faster sampling rate, typically in the order of 5MHz or higher.

One of the key challenges associated with these newer imaging techniques is achieving uniform excitation across the entire bandwidth spanned by the gradient for spatial encoding. Variations in the Flip Angle (α) can lead to changes in gradient direction, potentially resulting in image artifacts[132]. To mitigate the limitation of a limited Flip Angle, the application of RF excitation with amplitude modulation and frequency sweeps proves to be promising. Consequently, developments in creating RF pulses with short duration, high time resolution, rapid switching speed, variable rate digital filters, and ring down suppression form the foundation for future hardware advancements in clinical scanners. Notably, the ZTE technique demands dedicated, high-performance RF hardware[130] due to its minimal echo time. This is particularly critical in suppressing various shader artifacts associated with foam pads and plastics within RF coils, as these artifacts become significant during the detection of white matter, where enhanced contrast is essential.

## References

[1] P. C. Lauterbur, "Image Formation by Induced Local Interactions: Examples Employing Nuclear Magnetic Resonance," *Nature,* vol. 242, no. 5394, pp. 190-191, 1973/03 1973, doi: 10.1038/242190a0.

[2] P. Mansfield, "Multi-planar image formation using NMR spin echoes," *Journal of Physics C: Solid State Physics,* vol. 10, no. 3, pp. L55-L58, 1977/02/14 1977, doi: 10.1088/0022-3719/10/3/004.

[3] M. E. Ladd *et al.*, "Pros and cons of ultra-high-field MRI/MRS for human application," *Progress in Nuclear Magnetic Resonance Spectroscopy,* vol. 109, pp. 1-50, 2018/12 2018, doi: 10.1016/j.pnmrs.2018.06.001.

[4] K. Ugurbil *et al.*, "Imaging at high magnetic fields: initial experiences at 4 T," *Magn Reson Q,* vol. 9, no. 4, pp. 259-277, 1993.

[5] X. Zhang *et al.*, "High Resolution Imaging of the Human Head at 8 Tesla," in *Proceedings of ESMRMB annual meeting, Sevilla, Spain*, Sevilla, Spain, 1999, p. 44.

[6] K. Uğurbil *et al.*, "Magnetic Resonance Studies of Brain Function and Neurochemistry," *Annual Review of Biomedical Engineering,* vol. 2, no. 1, pp. 633-660, 2000/08 2000, doi: 10.1146/annurev.bioeng.2.1.633.

[7] X. Zhang, K. Ugurbil, and W. Chen, "Microstrip RF surface coil design for extremely high‐field MRI and spectroscopy," *Magnetic Resonance in Medicine,* vol. 46, no. 3, pp. 443-450, 2001/08/31 2001, doi: 10.1002/mrm.1212.

[8] X. Zhang, K. Ugurbil, and W. Chen, "Microstrip RF Surface Coils for Human MRI Studies at 7 Tesla," in *Proceedings of the 9th Annual Meeting of ISMRM, Glasgow, Scotland*, 2001, p. 1104.

[9] X. Zhang, K. Ugurbil, and W. Chen, "A New RF Volume Coil for Human MR Imaging and Spectroscopy at 4 Tesla," in *Proceedings of the 9th Annual Meeting of ISMRM, Glasgow, Scotland*, Glasgow, Scotland, 2001, p. 1103.

[10] X. Zhu, X. Zhang, S. Tang, S. Ogawa, K. Ugurbil, and W. Chen, "Probing fast neuronal interaction in the human ocular dominate columns based on fMRI BOLD response at 7 Tesla," in *Proceedings of the 9th Annual Meeting of ISMRM, Glasgow, Scotland*, 2001.

[11] X. Zhang, X. Zhu, H. Lei, Y. Zhang, and W. Chen, "A 400 MHz flexible MTL RF coil using the 2nd harmonic resonance for rat MR imaging at 9.4T," in *Proceedings of the 10th Annual Meeting of ISMRM, Honolulu, Hawaii*, 2002, p. 159.

[12] H. Qiao, X. Zhang, X.-H. Zhu, F. Du, and W. Chen, "In vivo 31P MRS of human brain at high/ultrahigh fields: a quantitative comparison of NMR detection sensitivity and spectral resolution between 4 T and 7 T," (in eng), *Magnetic resonance imaging,* vol. 24, no. 10, pp. 1281-1286, 2006, doi: 10.1016/j.mri.2006.08.002.

[13] X. Zhang, K. Ugurbil, and W. Chen, "A microstrip transmission line volume coil for human head MR imaging at 4T," *Journal of Magnetic Resonance,* vol. 161, no. 2, pp. 242-251, 2003/04 2003, doi: 10.1016/s1090-7807(03)00004-1.

[14] X. Zhang, X.-H. Zhu, R. Tian, Y. Zhang, H. Merkle, and W. Chen, "Measurement of arterial input function of 17O water tracer in rat carotid artery by using a region-defined (REDE) implanted vascular RF coil," *Magnetic Resonance Materials in Physics, Biology and Medicine,* vol. 16, no. 2, pp. 77-85, 2003/07 2003, doi: 10.1007/s10334-003-0013-9.

[15] X. Zhang, K. Ugurbil, and W. Chen, "Method and apparatus for magnetic resonance imaging and spectroscopy using microstrip transmission line coils," Patent Appl. 11224436, 2005.

[16] X. Zhang, K. Ugurbil, R. Sainati, and W. Chen, "An Inverted-Microstrip Resonator for Human Head Proton MR Imaging at 7 Tesla," *IEEE Transactions on Biomedical Engineering,* vol. 52, no. 3, pp. 495-504, 2005/03 2005, doi: 10.1109/tbme.2004.842968.

[17] X. Zhang, X. H. Zhu, and W. Chen, "Higher‐order harmonic transmission‐line RF coil design for MR applications," *Magnetic Resonance in Medicine,* vol. 53, no. 5, pp. 1234-1239, 2005/04/20 2005, doi: 10.1002/mrm.20462.

[18] H. Lei, X. H. Zhu, X. L. Zhang, K. Ugurbil, and W. Chen, "In vivo <sup>31</sup>P magnetic resonance spectroscopy of human brain at 7 T: An initial experience," *Magnetic Resonance in Medicine,* vol. 49, no. 2, pp. 199-205, 2003/01/22 2003, doi: 10.1002/mrm.10379.

[19] X. Zhang, K. Ugurbil, and W. Chen, "Method and apparatus for magnetic resonance imaging and spectroscopy using microstrip transmission line coils," (in English), Patent 7023209 Patent Appl. 09/974,184, 2006.
[20] X. Zhang and Z. Xie, "Method and apparatus for magnetic resonance imaging and spectroscopy using multiple-mode coils," Patent Appl. 12/990,541, 2009.
[21] X.-H. Zhu *et al.*, "Advanced In Vivo Heteronuclear MRS Approaches for Studying Brain Bioenergetics Driven by Mitochondria," (in eng), *Methods Mol Biol,* vol. 489, pp. 317-357, 2009, doi: 10.1007/978-1-59745-543-5_15.
[22] H. Dafni *et al.*, "Hyperpolarized 13C spectroscopic imaging informs on hypoxia-inducible factor-1 and myc activity downstream of platelet-derived growth factor receptor," (in eng), *Cancer Res,* vol. 70, no. 19, pp. 7400-7410, 2010, doi: 10.1158/0008-5472.CAN-10-0883.
[23] E. Moser, "Ultra-high-field magnetic resonance: Why and when?," (in eng), *World J Radiol,* vol. 2, no. 1, pp. 37-40, 2010, doi: 10.4329/wjr.v2.i1.37.
[24] Y. Pang and X. Zhang, "Interpolated Compressed Sensing MR Image Reconstruction using Neighboring Slice k-space Data," in *Proceedings of the 20th Annual Meeting of ISMRM, Melbourne, Australia*, 2012, p. 2275.
[25] Y. Pang and X. Zhang, "Interpolated compressed sensing for 2D multiple slice fast MR imaging," (in eng), *PLoS One,* vol. 8, no. 2, pp. e56098-e56098, 2013, doi: 10.1371/journal.pone.0056098.
[26] X. Zhang *et al.*, "Human extremity imaging using microstrip resonators at 7T," in *Proceedings of the 21st Annual Meeting of ISMRM, Salt Lake City, USA*, 2013, p. 1675.
[27] Y. X. Wang, G. G. Lo, J. Yuan, P. E. Larson, and X. Zhang, "Magnetic resonance imaging for lung cancer screen," *J Thorac Dis,* vol. 6, no. 9, pp. 1340-8, Sep 2014, doi: 10.3978/j.issn.2072-1439.2014.08.43.
[28] E. Milshteyn and X. Zhang, "The Need and Initial Practice of Parallel Imaging and Compressed Sensing in Hyperpolarized (13)C MRI in vivo," (in eng), *OMICS J Radiol,* vol. 4, no. 4, p. e133, 2015, doi: 10.4172/2167-7964.1000e133.
[29] X. Zhang, "Sensitivity enhancement of traveling wave MRI using free local resonators: an experimental demonstration," (in eng), *Quant Imaging Med Surg,* vol. 7, no. 2, pp. 170-176, 2017, doi: 10.21037/qims.2017.02.10.
[30] Z. Wei *et al.*, "5T magnetic resonance imaging: radio frequency hardware and initial brain imaging," *Quant Imaging Med Surg,* vol. 13, no. 5, pp. 3222-3240, May 1 2023, doi: 10.21037/qims-22-945.
[31] X. Zhang, L. DelaBarre, K. Payne, M. Waks, G. Adriany, and K. Ugurbil, "A Wrap-on Decoupled Coaxial Transmission Line (CTL) Transceive Array for Head MR Imaging at 10.5T," *Proc Int Soc Magn Reson Med Sci Meet Exhib Int Soc Magn Reson Med Sci Meet Exhib,* vol. 31, Jun 2023. [Online]. Available: https://www.ncbi.nlm.nih.gov/pubmed/37609489.
[32] X. Zhang, M. Waks, L. DelaBarre, K. Payne, K. Ugurbil, and G. Adriany, "Design and Test of a Flexible Two-row CTL Array and Its Detunable Resonant Elements for 10.5 T MR Imaging," *Proc Intl Soc Mag Reson Med,* p. 4593, 2023.
[33] K. Payne, A. A. Bhosale, and X. Zhang, "Double cross magnetic wall decoupling for quadrature transceiver RF array coils using common-mode differential-mode

resonators," *J Magn Reson,* vol. 353, p. 107498, Aug 2023, doi: 10.1016/j.jmr.2023.107498.
[34] D. I. Hoult and R. E. Richards, "The signal-to-noise ratio of the nuclear magnetic resonance experiment," *Journal of Magnetic Resonance (1969),* vol. 24, no. 1, pp. 71-85, 1976/10 1976, doi: 10.1016/0022-2364(76)90233-x.
[35] J. J. H. Ackerman, T. H. Grove, G. G. Wong, D. G. Gadian, and G. K. Radda, "Mapping of metabolites in whole animals by 31P NMR using surface coils," *Nature,* vol. 283, no. 5743, pp. 167-170, 1980/01 1980, doi: 10.1038/283167a0.
[36] C. N. Chen, V. J. Sank, S. M. Cohen, and D. I. Hoult, "The field dependence of NMR imaging. I. Laboratory assessment of signal‐to‐noise ratio and power deposition," *Magnetic Resonance in Medicine,* vol. 3, no. 5, pp. 722-729, 1986/10 1986, doi: 10.1002/mrm.1910030508.
[37] D. I. Hoult, "Sensitivity and Power Deposition in a High-Field Imaging Experiment," *Journal of Magnetic Resonance Imaging,* vol. 12, no. 1, pp. 46-67, 2000, doi: 10.1002/1522-2586(200007)12:1<46::aid-jmri6>3.0.co;2-d.
[38] X. Zhang, "Experimental Design of Transmission Line Volume RF coil for MR Imaging at 8T," in *Proceedings of the 8th Annual Meeting of ISMRM, Denver, USA*, 2000, p. 150.
[39] X. Zhang, "Novel radio frequency resonators for in vivo magnetic resonance imaging and spectroscopy at very high magnetic fields," Ph.D., Biomedical Engineering, University of Minnesota, 2002.
[40] X.-H. Zhu *et al.*, "Development of (17)O NMR approach for fast imaging of cerebral metabolic rate of oxygen in rat brain at high field," (in eng), *Proc Natl Acad Sci U S A,* vol. 99, no. 20, pp. 13194-13199, 2002, doi: 10.1073/pnas.202471399.
[41] C. Wang and X. Zhang, "Evaluation of B1+ and E field of RF Resonator with High Dielectric Insert," in *Proceedings of the 17th Annual Meeting of ISMRM, Honolulu, Hawaii*, 2009, p. 3054.
[42] B. Wu, Y. Li, C. Wang, D. B. Vigneron, and X. Zhang, "Multi-reception strategy with improved SNR for multichannel MR imaging," (in eng), *PLoS One,* vol. 7, no. 8, pp. e42237-e42237, 2012, doi: 10.1371/journal.pone.0042237.
[43] J. Kurhanewicz *et al.*, "Hyperpolarized (13)C MRI: Path to Clinical Translation in Oncology," (in eng), *Neoplasia,* vol. 21, no. 1, pp. 1-16, 2019, doi: 10.1016/j.neo.2018.09.006.
[44] J. T. Vaughan *et al.*, "7T vs. 4T: RF power, homogeneity, and signal‐to‐noise comparison in head images," *Magnetic Resonance in Medicine,* vol. 46, no. 1, pp. 24-30, 2001/07 2001, doi: 10.1002/mrm.1156.
[45] R. Krug *et al.*, "Ultrashort echo time MRI of cortical bone at 7 tesla field strength: a feasibility study," *J Magn Reson Imaging,* vol. 34, no. 3, pp. 691-5, Sep 2011, doi: 10.1002/jmri.22648.
[46] D. I. Hoult, "The principle of reciprocity in signal strength calculations?A mathematical guide," *Concepts in Magnetic Resonance,* vol. 12, no. 4, pp. 173-187, 2000, doi: 10.1002/1099-0534(2000)12:4<173::aid-cmr1>3.0.co;2-q.
[47] C. M. Collins *et al.*, "Different excitation and reception distributions with a single‐loop transmit‐receive surface coil near a head‐sized spherical phantom at 300 MHz,"

*Magnetic Resonance in Medicine,* vol. 47, no. 5, pp. 1026-1028, 2002/04/22 2002, doi: 10.1002/mrm.10153.
[48] Q. X. Yang *et al.*, "Analysis of wave behavior in lossy dielectric samples at high field," *Magnetic Resonance in Medicine,* vol. 47, no. 5, pp. 982-989, 2002/04/23 2002, doi: 10.1002/mrm.10137.
[49] G. Adriany *et al.*, "Transmit and receive transmission line arrays for 7 Tesla parallel imaging," *Magnetic Resonance in Medicine,* vol. 53, no. 2, pp. 434-445, 2005/01/27 2005, doi: 10.1002/mrm.20321.
[50] X. Zhang, A. R. Burr, X. Zhu, G. Adriany, K. Ugurbil, and W. Chen, "A Dual-tuned Microstrip Volume Coil Array for Human Head parallel 1H/31P MRI/MRS at 7T," in *the 11th Scientific Meeting and Exhibition of ISMRM*, Miami, Florida, 2005, p. 896.
[51] Q. X. Yang *et al.*, "Manipulation of image intensity distribution at 7.0 T: Passive RF shimming and focusing with dielectric materials," *Journal of Magnetic Resonance Imaging,* vol. 24, no. 1, pp. 197-202, 2006/06/05 2006, doi: 10.1002/jmri.20603.
[52] B. Wu, X. Zhang, P. Qu, and G. X. Shen, "Capacitively decoupled tunable loop microstrip (TLM) array at 7 T," *Magnetic Resonance Imaging,* vol. 25, no. 3, pp. 418-424, 2007/04 2007, doi: 10.1016/j.mri.2006.09.031.
[53] Z. Xie and X. Zhang, "A novel decoupling technique for non-overlapped microstrip array coil at 7T MR imaging," in *the 16th Annual Meeting of ISMRM*, Toronto, Canada, P. o. t. t. A. M. o. ISMRM, Ed., May 3 -9 2008, p. 1068.
[54] Z. Xie and X. Zhang, "An 8-channel non-overlapped spinal cord array coil for 7T MR imaging," in *the 16th Annual Meeting of ISMRM*, Toronto, Canada, P. o. t. t. A. M. o. ISMRM, Ed., May 3 -9 2008, p. 2974.
[55] Z. Xie and X. Zhang, "An 8-channel microstrip array coil for mouse parallel MR imaging at 7T by using magnetic wall decoupling technique," in *the 16th Annual Meeting of ISMRM*, Toronto, Canada, P. o. t. t. A. M. o. ISMRM, Ed., May 3 -9 2008, p. 2973.
[56] Y. Li, Y. Pang, and X. Zhang, "Common-mode differential-mode (CMDM) method for quadrature transmit/receive surface coil for ultrahigh field MRI," in *the 19th Annual Meeting of ISMRM*, Montreal, Canada, 2011.
[57] Y. Li, Z. Xie, Y. Pang, D. Vigneron, and X. Zhang, "ICE decoupling technique for RF coil array designs," (in eng), *Medical physics,* vol. 38, no. 7, pp. 4086-4093, 2011, doi: 10.1118/1.3598112.
[58] J. Lu, Y. Pang, C. Wang, B. Wu, D. B. Vigneron, and X. Zhang, "Evaluation of Common RF Coil Setups for MR Imaging at Ultrahigh Magnetic Field: A Numerical Study," (in eng), *Int Symp Appl Sci Biomed Commun Technol,* vol. 2011, p. 70, 2011, doi: 10.1145/2093698.2093768.
[59] Y. Pang, B. Wu, C. Wang, D. B. Vigneron, and X. Zhang, "Numerical Analysis of Human Sample Effect on RF Penetration and Liver MR Imaging at Ultrahigh Field," (in eng), *Concepts Magn Reson Part B Magn Reson Eng,* vol. 39B, no. 4, pp. 206-216, 2011, doi: 10.1002/cmr.b.20209.
[60] Y. Pang and X. Zhang, "Precompensation for mutual coupling between array elements in parallel excitation," *Quant Imaging Med Surg,* vol. 1, no. 1, pp. 4-10, Dec 2011. [Online]. Available:

http://www.ncbi.nlm.nih.gov/entrez/query.fcgi?cmd=Retrieve&db=PubMed&dopt=Citation&list_uids=23243630
[61] Y. Pang, X. Zhang, Z. Xie, C. Wang, and D. B. Vigneron, "Common-mode differential-mode (CMDM) method for double-nuclear MR signal excitation and reception at ultrahigh fields," (in eng), *IEEE Trans Med Imaging,* vol. 30, no. 11, pp. 1965-1973, 2011, doi: 10.1109/TMI.2011.2160192.
[62] K. Subburaj *et al.*, "A Flexible Microstrip Transceiver Coil for Imaging Flexed Human Knee Joints at 7 Tesla," in *Proceedings of the 19th Annual Meeting of ISMRM, Montreal, Canada*, 2011, p. 3821.
[63] Y. Pang, D. B. Vigneron, and X. Zhang, "Parallel traveling-wave MRI: a feasibility study," (in eng), *Magnetic resonance in medicine,* vol. 67, no. 4, pp. 965-978, 2012, doi: 10.1002/mrm.23073.
[64] Y. Pang *et al.*, "A dual-tuned quadrature volume coil with mixed λ/2 and λ/4 microstrip resonators for multinuclear MRSI at 7 T," (in eng), *Magnetic resonance imaging,* vol. 30, no. 2, pp. 290-298, 2012, doi: 10.1016/j.mri.2011.09.022.
[65] C. Wang *et al.*, "A practical multinuclear transceiver volume coil for in vivo MRI/MRS at 7 T," (in eng), *Magnetic resonance imaging,* vol. 30, no. 1, pp. 78-84, 2012, doi: 10.1016/j.mri.2011.08.007.
[66] Y. Li, C. Wang, B. Yu, D. Vigneron, W. Chen, and X. Zhang, "Image homogenization using pre-emphasis method for high field MRI," *Quant Imaging Med Surg,* vol. 3, no. 4, pp. 217-23, Aug 2013. [Online]. Available: http://www.ncbi.nlm.nih.gov/entrez/query.fcgi?cmd=Retrieve&db=PubMed&dopt=Citation&list_uids=24040618
[67] X. Hu, X. Chen, X. Liu, H. Zheng, Y. Li, and X. Zhang, "Parallel imaging performance investigation of an 8-channel common-mode differential-mode (CMDM) planar array for 7T MRI," *Quant Imaging Med Surg,* vol. 4, no. 1, pp. 33-42, Feb 2014. [Online]. Available: http://www.ncbi.nlm.nih.gov/entrez/query.fcgi?cmd=Retrieve&db=PubMed&dopt=Citation&list_uids=24649433
[68] Y. Li, B. Yu, Y. Pang, D. B. Vigneron, and X. Zhang, "Planar quadrature RF transceiver design using common-mode differential-mode (CMDM) transmission line method for 7T MR imaging," (in eng), *PLoS One,* vol. 8, no. 11, pp. e80428-e80428, 2013, doi: 10.1371/journal.pone.0080428.
[69] Y. Pang, B. Wu, X. Jiang, D. B. Vigneron, and X. Zhang, "Tilted microstrip phased arrays with improved electromagnetic decoupling for ultrahigh-field magnetic resonance imaging," (in eng), *Medicine (Baltimore),* vol. 93, no. 28, pp. e311-e311, 2014, doi: 10.1097/MD.0000000000000311.
[70] X. Yan, R. Xue, and X. Zhang, "A monopole/loop dual-tuned RF coil for ultrahigh field MRI," *Quant Imaging Med Surg,* vol. 4, no. 4, pp. 225-31, Aug 2014. [Online]. Available: http://www.ncbi.nlm.nih.gov/entrez/query.fcgi?cmd=Retrieve&db=PubMed&dopt=Citation&list_uids=25202657
[71] X. Yan, L. Wei, R. Xue, and X. Zhang, "Hybrid monopole/loop coil array for human head MR imaging at 7T," (in eng), *Appl Magn Reson,* vol. 46, no. 5, pp. 541-550, 2015, doi: 10.1007/s00723-015-0656-5.

[72] X. Yan and X. Zhang, "Decoupling and matching network for monopole antenna arrays in ultrahigh field MRI," *Quant Imaging Med Surg,* vol. 5, no. 4, pp. 546-51, Aug 2015, doi: 10.3978/j.issn.2223-4292.2015.07.06.
[73] X. Yan, L. Wei, S. Chu, R. Xue, and X. Zhang, "Eight-Channel Monopole Array Using ICE Decoupling for Human Head MR Imaging at 7 T," (in eng), *Appl Magn Reson,* vol. 47, no. 5, pp. 527-538, 2016, doi: 10.1007/s00723-016-0775-7.
[74] X. Yan, R. Xue, and X. Zhang, "Closely-spaced double-row microstrip RF arrays for parallel MR imaging at ultrahigh fields," (in eng), *Appl Magn Reson,* vol. 46, no. 11, pp. 1239-1248, 2015, doi: 10.1007/s00723-015-0712-1.
[75] X. Yan, Z. Cao, and X. Zhang, "Simulation verification of SNR and parallel imaging improvements by ICE-decoupled loop array in MRI," (in eng), *Appl Magn Reson,* vol. 47, no. 4, pp. 395-403, 2016, doi: 10.1007/s00723-016-0764-x.
[76] B. Wu *et al.*, "Shielded microstrip array for 7T human MR imaging," (in eng), *IEEE Trans Med Imaging,* vol. 29, no. 1, pp. 179-184, 2010, doi: 10.1109/TMI.2009.2033597.
[77] B. Wu *et al.*, "7T human spine imaging arrays with adjustable inductive decoupling," (in eng), *IEEE Trans Biomed Eng,* vol. 57, no. 2, pp. 397-403, 2010, doi: 10.1109/TBME.2009.2030170.
[78] B. Wu *et al.*, "Multi-channel microstrip transceiver arrays using harmonics for high field MR imaging in humans," (in eng), *IEEE Trans Med Imaging,* vol. 31, no. 2, pp. 183-191, 2012, doi: 10.1109/TMI.2011.2166273.
[79] B. Wu *et al.*, "Flexible transceiver array for ultrahigh field human MR imaging," (in eng), *Magnetic resonance in medicine,* vol. 68, no. 4, pp. 1332-1338, 2012, doi: 10.1002/mrm.24121.
[80] K. Payne, L. L. Ying, and X. Zhang, "Hairpin RF resonators for MR imaging transceiver arrays with high inter-channel isolation and B(1) efficiency at ultrahigh field 7 T," *J Magn Reson,* vol. 345, p. 107321, Dec 2022, doi: 10.1016/j.jmr.2022.107321.
[81] K. Payne, Y. Zhao, A. A. Bhosale, and X. Zhang, "Dual-tuned Coaxial-transmission-line RF coils for Hyperpolarized (13)C and Deuterium (2)H Metabolic MRS Imaging at Ultrahigh Fields," *IEEE Trans Biomed Eng,* vol. PP, Dec 12 2023, doi: 10.1109/TBME.2023.3341760.
[82] C. Li, F. W. Wehrli, and University of Pennsylvania. Bioengineering., *Magnetic resonance imaging of short-T2 tissues with applications for quantifying cortical bone water and myelin*. p. 1 online resource (145 pages).
[83] Z. Wu *et al.*, "Assessment of ultrashort echo time (UTE) T(2)* mapping at 3T for the whole knee: repeatability, the effects of fat suppression, and knee position," *Quant Imaging Med Surg,* vol. 13, no. 12, pp. 7893-7909, Dec 1 2023, doi: 10.21037/qims-23-459.
[84] J. M. Pauly, *SMRI,* p. 330, 1992.
[85] M. Weiger and K. P. Pruessmann, "MRI with Zero Echo Time," *eMagRes,* vol. 1, pp. 311-322, 2012, doi: DOI 10.1002/9780470034590.emrstm1292.
[86] P. E. Z. Larson *et al.*, "Ultrashort echo time and zero echo time MRI at 7T," (in eng), *MAGMA,* vol. 29, no. 3, pp. 359-370, 2016, doi: 10.1007/s10334-015-0509-0.
[87] M. D. Robson and G. M. Bydder, "Clinical ultrashort echo time imaging of bone and other connective tissues," *NMR Biomed,* vol. 19, no. 7, pp. 765-80, Nov 2006, doi: 10.1002/nbm.1100.

[88] G. E. Gold, E. Han, J. Stainsby, G. Wright, J. Brittain, and C. Beaulieu, "Musculoskeletal MRI at 3.0 T: relaxation times and image contrast," *AJR Am J Roentgenol,* vol. 183, no. 2, pp. 343-51, Aug 2004, doi: 10.2214/ajr.183.2.1830343.

[89] A. Pai, X. Li, and S. Majumdar, "A comparative study at 3 T of sequence dependence of T2 quantitation in the knee," *Magn Reson Imaging,* vol. 26, no. 9, pp. 1215-20, Nov 2008, doi: 10.1016/j.mri.2008.02.017.

[90] J. Desrochers, A. Yung, D. Stockton, and D. Wilson, "Depth-dependent changes in cartilage T2 under compressive strain: a 7T MRI study on human knee cartilage," *Osteoarthritis Cartilage,* vol. 28, no. 9, pp. 1276-1285, Sep 2020, doi: 10.1016/j.joca.2020.05.012.

[91] G. H. Filho *et al.*, "Quantitative characterization of the Achilles tendon in cadaveric specimens: T1 and T2* measurements using ultrashort-TE MRI at 3 T," *AJR Am J Roentgenol,* vol. 192, no. 3, pp. W117-24, Mar 2009, doi: 10.2214/AJR.07.3990.

[92] M. Han, P. E. Larson, J. Liu, and R. Krug, "Depiction of achilles tendon microstructure in vivo using high-resolution 3-dimensional ultrashort echo-time magnetic resonance imaging at 7 T," *Invest Radiol,* vol. 49, no. 5, pp. 339-45, May 2014, doi: 10.1097/RLI.0000000000000025.

[93] M. C. Wurnig *et al.*, "Characterization of trabecular bone density with ultra-short echo-time MRI at 1.5, 3.0 and 7.0 T--comparison with micro-computed tomography," *NMR Biomed,* vol. 27, no. 10, pp. 1159-66, Oct 2014, doi: 10.1002/nbm.3169.

[94] M. Okuda *et al.*, "Quantitative differentiation of tendon and ligament using magnetic resonance imaging ultrashort echo time T2* mapping of normal knee joint," *Acta Radiol,* p. 2841851211043834, Sep 24 2021, doi: 10.1177/02841851211043834.

[95] U. D. Monu, C. D. Jordan, B. L. Samuelson, B. A. Hargreaves, G. E. Gold, and E. J. McWalter, "Cluster analysis of quantitative MRI T2 and T1rho relaxation times of cartilage identifies differences between healthy and ACL-injured individuals at 3T," *Osteoarthritis Cartilage,* vol. 25, no. 4, pp. 513-520, Apr 2017, doi: 10.1016/j.joca.2016.09.015.

[96] I. Blystad *et al.*, "Quantitative MRI for Analysis of Active Multiple Sclerosis Lesions without Gadolinium-Based Contrast Agent," *AJNR Am J Neuroradiol,* vol. 37, no. 1, pp. 94-100, Jan 2016, doi: 10.3174/ajnr.A4501.

[97] M. Neema *et al.*, "3 T MRI relaxometry detects T2 prolongation in the cerebral normal-appearing white matter in multiple sclerosis," *Neuroimage,* vol. 46, no. 3, pp. 633-41, Jul 1 2009, doi: 10.1016/j.neuroimage.2009.03.001.

[98] C. Mainero *et al.*, "A gradient in cortical pathology in multiple sclerosis by in vivo quantitative 7 T imaging," *Brain,* vol. 138, no. Pt 4, pp. 932-45, Apr 2015, doi: 10.1093/brain/awv011.

[99] T. Heye *et al.*, "MR relaxometry of the liver: significant elevation of T1 relaxation time in patients with liver cirrhosis," *Eur Radiol,* vol. 22, no. 6, pp. 1224-32, Jun 2012, doi: 10.1007/s00330-012-2378-5.

[100] V. C. Obmann *et al.*, "Liver MR relaxometry at 3T - segmental normal T1 and T2* values in patients without focal or diffuse liver disease and in patients with increased liver fat and elevated liver stiffness," *Sci Rep,* vol. 9, no. 1, p. 8106, May 30 2019, doi: 10.1038/s41598-019-44377-y.

[101] C. D. Jordan, M. Saranathan, N. K. Bangerter, B. A. Hargreaves, and G. E. Gold, "Musculoskeletal MRI at 3.0 T and 7.0 T: a comparison of relaxation times and image contrast," *Eur J Radiol,* vol. 82, no. 5, pp. 734-9, May 2013, doi: 10.1016/j.ejrad.2011.09.021.

[102] C. A. Hanson, A. Kamath, M. Gottbrecht, S. Ibrahim, and M. Salerno, "T2 Relaxation Times at Cardiac MRI in Healthy Adults: A Systematic Review and Meta-Analysis," *Radiology,* vol. 297, no. 2, pp. 344-351, Nov 2020, doi: 10.1148/radiol.2020200989.

[103] T. Huelnhagen *et al.*, "Myocardial Effective Transverse Relaxation Time T 2(*) is Elevated in Hypertrophic Cardiomyopathy: A 7.0 T Magnetic Resonance Imaging Study," *Sci Rep,* vol. 8, no. 1, p. 3974, Mar 5 2018, doi: 10.1038/s41598-018-22439-x.

[104] A. M. Peters *et al.*, "T2* measurements in human brain at 1.5, 3 and 7 T," *Magn Reson Imaging,* vol. 25, no. 6, pp. 748-53, Jul 2007, doi: 10.1016/j.mri.2007.02.014.

[105] J. P. Wansapura, S. K. Holland, R. S. Dunn, and W. S. Ball Jr., "NMR relaxation times in the human brain at 3.0 tesla," *Journal of Magnetic Resonance Imaging,* vol. 9, no. 4, pp. 531-538, 1999, doi: https://doi.org/10.1002/(SICI)1522-2586(199904)9:4<531::AID-JMRI4>3.0.CO;2-L.

[106] N. Gelman *et al.*, "MR imaging of human brain at 3.0 T: preliminary report on transverse relaxation rates and relation to estimated iron content," *Radiology,* vol. 210, no. 3, pp. 759-67, Mar 1999, doi: 10.1148/radiology.210.3.r99fe41759.

[107] J. L. Zhang *et al.*, "Reproducibility of R2* and R2 measurements in human kidneys," in *Proc Int Soc Magn Reson Med*, 2011, vol. 19, p. 2954.

[108] X. Li, P. J. Bolan, K. Ugurbil, and G. J. Metzger, "Measuring renal tissue relaxation times at 7 T," *NMR Biomed,* vol. 28, no. 1, pp. 63-9, Jan 2015, doi: 10.1002/nbm.3195.

[109] J. T. Bushberg and J. M. Boone, *The Essential Physics of Medical Imaging*. Wolters Kluwer Health, 2011.

[110] A. Daoust *et al.*, "Transverse relaxation of cerebrospinal fluid depends on glucose concentration," *Magn Reson Imaging,* vol. 44, pp. 72-81, Dec 2017, doi: 10.1016/j.mri.2017.08.001.

[111] M. Barth and E. Moser, "Proton NMR relaxation times of human blood samples at 1.5 T and implications for functional MRI," *Cell Mol Biol (Noisy-le-grand),* vol. 43, no. 5, pp. 783-91, Jul 1997. [Online]. Available: https://www.ncbi.nlm.nih.gov/pubmed/9298600.

[112] F. H. P. van Leeuwen *et al.*, "Detecting low blood concentrations in joints using T1 and T2 mapping at 1.5, 3, and 7 T: an in vitro study," *Eur Radiol Exp,* vol. 5, no. 1, p. 51, Dec 2 2021, doi: 10.1186/s41747-021-00251-z.

[113] L. C. Krishnamurthy, P. Liu, F. Xu, J. Uh, I. Dimitrov, and H. Lu, "Dependence of blood T(2) on oxygenation at 7 T: in vitro calibration and in vivo application," *Magn Reson Med,* vol. 71, no. 6, pp. 2035-42, Jun 2014, doi: 10.1002/mrm.24868.

[114] A. L. Lin, Q. Qin, X. Zhao, and T. Q. Duong, "Blood longitudinal (T1) and transverse (T2) relaxation time constants at 11.7 Tesla," *MAGMA,* vol. 25, no. 3, pp. 245-9, Jun 2012, doi: 10.1007/s10334-011-0287-2.

[115] D. J. Tyler, M. D. Robson, R. M. Henkelman, I. R. Young, and G. M. Bydder, "Magnetic resonance imaging with ultrashort TE (UTE) PULSE sequences: technical considerations," *J Magn Reson Imaging,* vol. 25, no. 2, pp. 279-89, Feb 2007, doi: 10.1002/jmri.20851.

[116] Y.-J. Ma, S. Jerban, H. Jang, D. Chang, E. Y. Chang, and J. Du, "Quantitative Ultrashort Echo Time (UTE) Magnetic Resonance Imaging of Bone: An Update," (in eng), *Front Endocrinol (Lausanne),* vol. 11, pp. 567417-567417, 2020, doi: 10.3389/fendo.2020.567417.

[117] H. T. Fabich, M. Benning, A. J. Sederman, and D. J. Holland, "Ultrashort echo time (UTE) imaging using gradient pre-equalization and compressed sensing," *Journal of Magnetic Resonance,* vol. 245, pp. 116-124, 2014/08 2014, doi: 10.1016/j.jmr.2014.06.015.

[118] C. Li, J. F. Magland, A. C. Seifert, and F. W. Wehrli, "Correction of excitation profile in Zero Echo Time (ZTE) imaging using quadratic phase-modulated RF pulse excitation and iterative reconstruction," (in eng), *IEEE Trans Med Imaging,* vol. 33, no. 4, pp. 961-969, 2014, doi: 10.1109/TMI.2014.2300500.

[119] V. Juras *et al.*, "Regional variations of <i>T</i><sub>2</sub>* in healthy and pathologic achilles tendon in vivo at 7 Tesla: Preliminary results," *Magnetic Resonance in Medicine,* vol. 68, no. 5, pp. 1607-1613, 2012/01/03 2012, doi: 10.1002/mrm.24136.

[120] Y. Wu *et al.*, "Water- and fat-suppressed proton projection MRI (WASPI) of rat femur bone," *Magn Reson Med,* vol. 57, no. 3, pp. 554-67, Mar 2007, doi: 10.1002/mrm.21174.

[121] X. Zhao, H. K. Song, A. C. Seifert, C. Li, and F. W. Wehrli, "Feasibility of assessing bone matrix and mineral properties in vivo by combined solid-state 1H and 31P MRI," (in eng), *PLoS One,* vol. 12, no. 3, pp. e0173995-e0173995, 2017, doi: 10.1371/journal.pone.0173995.

[122] J. Du, A. M. Takahashi, W. C. Bae, C. B. Chung, and G. M. Bydder, "Dual inversion recovery, ultrashort echo time (DIR UTE) imaging: creating high contrast for short-T(2) species," (in eng), *Magnetic resonance in medicine,* vol. 63, no. 2, pp. 447-455, 2010, doi: 10.1002/mrm.22257.

[123] D. Idiyatullin, C. Corum, J. Y. Park, and M. Garwood, "Fast and quiet MRI using a swept radiofrequency," *J Magn Reson,* vol. 181, no. 2, pp. 342-9, Aug 2006, doi: 10.1016/j.jmr.2006.05.014.

[124] D. Idiyatullin, C. A. Corum, and M. Garwood, "Multi-Band-SWIFT," (in eng), *J Magn Reson,* vol. 251, pp. 19-25, 2015, doi: 10.1016/j.jmr.2014.11.014.

[125] K. H. Herrmann, M. Kramer, and J. R. Reichenbach, "Time Efficient 3D Radial UTE Sampling with Fully Automatic Delay Compensation on a Clinical 3T MR Scanner," *PLoS One,* vol. 11, no. 3, p. e0150371, 2016, doi: 10.1371/journal.pone.0150371.

[126] J. Du, M. Bydder, A. M. Takahashi, and C. B. Chung, "Two-dimensional ultrashort echo time imaging using a spiral trajectory," *Magnetic Resonance Imaging,* vol. 26, no. 3, pp. 304-312, 2008/04 2008, doi: 10.1016/j.mri.2007.08.005.

[127] L. Wan *et al.*, "Fast quantitative three-dimensional ultrashort echo time (UTE) Cones magnetic resonance imaging of major tissues in the knee joint using extended sprial sampling," (in eng), *NMR in biomedicine,* vol. 33, no. 10, pp. e4376-e4376, 2020, doi: 10.1002/nbm.4376.

[128] Y. Qian, T. Zhao, Y.-K. Hue, T. S. Ibrahim, and F. E. Boada, "High-resolution spiral imaging on a whole-body 7T scanner with minimized image blurring," *Magnetic Resonance in Medicine,* vol. 63, no. 3, pp. 543-552, 2010/02/09 2010, doi: 10.1002/mrm.22215.

[129] C. Li, J. F. Magland, H. S. Rad, H. K. Song, and F. W. Wehrli, "Comparison of optimized soft-tissue suppression schemes for ultrashort echo time MRI," (in eng), *Magnetic resonance in medicine,* vol. 68, no. 3, pp. 680-689, 2012, doi: 10.1002/mrm.23267.
[130] M. Weiger, D. O. Brunner, B. E. Dietrich, C. F. Müller, and K. P. Pruessmann, "ZTE imaging in humans," *Magnetic Resonance in Medicine,* vol. 70, no. 2, pp. 328-332, 2013/06/14 2013, doi: 10.1002/mrm.24816.
[131] R. E. Breighner, Y. Endo, G. P. Konin, L. V. Gulotta, M. F. Koff, and H. G. Potter, "Technical Developments: Zero Echo Time Imaging of the Shoulder: Enhanced Osseous Detail by Using MR Imaging," *Radiology,* vol. 286, no. 3, pp. 960-966, Mar 2018, doi: 10.1148/radiol.2017170906.
[132] K. Schieban, M. Weiger, F. Hennel, A. Boss, and K. P. Pruessmann, "ZTE imaging with enhanced flip angle using modulated excitation," *Magn Reson Med,* vol. 74, no. 3, pp. 684-93, Sep 2015, doi: 10.1002/mrm.25464.
[133] E. Y. Chang, J. Du, and C. B. Chung, "UTE imaging in the musculoskeletal system," (in eng), *J Magn Reson Imaging,* vol. 41, no. 4, pp. 870-883, 2015, doi: 10.1002/jmri.24713.
[134] J. Yang *et al.*, "Quantitative ultrashort echo time magnetization transfer (UTE-MT) for diagnosis of early cartilage degeneration: comparison with UTE-T2* and T2 mapping," (in eng), *Quant Imaging Med Surg,* vol. 10, no. 1, pp. 171-183, 2020, doi: 10.21037/qims.2019.12.04.
[135] G. Dournes *et al.*, "The Clinical Use of Lung MRI in Cystic Fibrosis: What, Now, How?," (in eng), *Chest,* vol. 159, no. 6, pp. 2205-2217, 2021, doi: 10.1016/j.chest.2020.12.008.
[136] K. W. Stock, Q. Chen, H. Hatabu, and R. R. Edelman, "Magnetic resonance T2∗ measurements of the normal human lung in vivo with ultra-short echo times," *Magnetic Resonance Imaging,* vol. 17, no. 7, pp. 997-1000, 1999/09 1999, doi: 10.1016/s0730-725x(99)00047-8.
[137] K. M. Johnson, S. B. Fain, M. L. Schiebler, and S. Nagle, "Optimized 3D ultrashort echo time pulmonary MRI," (in eng), *Magnetic resonance in medicine,* vol. 70, no. 5, pp. 1241-1250, 2013, doi: 10.1002/mrm.24570.
[138] D. O. Kuethe, A. Caprihan, E. Fukushima, and R. A. Waggoner, "Imaging lungs using inert fluorinated gases," *Magnetic Resonance in Medicine,* vol. 39, no. 1, pp. 85-88, 1998/01 1998, doi: 10.1002/mrm.1910390114.
[139] M. Weiger *et al.*, "Rapid and robust pulmonary proton ZTE imaging in the mouse," *NMR in Biomedicine,* vol. 27, no. 9, pp. 1129-1134, 2014/07/26 2014, doi: 10.1002/nbm.3161.
[140] J. Guo, X. Cao, Z. I. Cleveland, and J. C. Woods, "Murine pulmonary imaging at 7T: T2* and T(1) with anisotropic UTE," (in eng), *Magnetic resonance in medicine,* vol. 79, no. 4, pp. 2254-2264, 2018, doi: 10.1002/mrm.26872.
[141] N. Kobayashi, D. Idiyatullin, C. Corum, J. Weber, M. Garwood, and D. Sachdev, "SWIFT MRI enhances detection of breast cancer metastasis to the lung," *Magn Reson Med,* vol. 73, no. 5, pp. 1812-9, May 2015, doi: 10.1002/mrm.25301.
[142] J. B. Hövener *et al.*, "Dental MRI: Imaging of soft and solid components without ionizing radiation," *Journal of Magnetic Resonance Imaging,* vol. 36, no. 4, pp. 841-846, 2012/06/15 2012, doi: 10.1002/jmri.23712.
[143] S. Mastrogiacomo, W. Dou, J. A. Jansen, and X. F. Walboomers, "Magnetic Resonance Imaging of Hard Tissues and Hard Tissue Engineered Bio-substitutes," *Molecular*

*Imaging and Biology,* vol. 21, no. 6, pp. 1003-1019, 2019/11/07 2019, doi: 10.1007/s11307-019-01345-2.

[144] D. Idiyatullin, C. Corum, S. Moeller, H. S. Prasad, M. Garwood, and D. R. Nixdorf, "Dental magnetic resonance imaging: making the invisible visible," (in eng), *J Endod,* vol. 37, no. 6, pp. 745-752, 2011, doi: 10.1016/j.joen.2011.02.022.

[145] M. J. Wilhelm *et al.*, "Direct magnetic resonance detection of myelin and prospects for quantitative imaging of myelin density," (in eng), *Proc Natl Acad Sci U S A,* vol. 109, no. 24, pp. 9605-9610, 2012, doi: 10.1073/pnas.1115107109.

[146] H. Jang *et al.*, "Inversion Recovery Ultrashort TE MR Imaging of Myelin is Significantly Correlated with Disability in Patients with Multiple Sclerosis," (in eng), *AJNR Am J Neuroradiol,* vol. 42, no. 5, pp. 868-874, 2021, doi: 10.3174/ajnr.A7006.

[147] A. Inoue *et al.*, "Clinical utility of new three-dimensional model using a zero-echo-time sequence in endoscopic endonasal transsphenoidal surgery," *Clinical Neurology and Neurosurgery,* vol. 190, p. 105743, 2020/03 2020, doi: 10.1016/j.clineuro.2020.105743.

[148] R. A. Horch, J. C. Gore, and M. D. Does, "Origins of the ultrashort-T2 1H NMR signals in myelinated nerve: a direct measure of myelin content?," (in eng), *Magnetic resonance in medicine,* vol. 66, no. 1, pp. 24-31, 2011, doi: 10.1002/mrm.22980.

[149] F. Nelson, A. Poonawalla, P. Hou, J. S. Wolinsky, and P. A. Narayana, "3D MPRAGE improves classification of cortical lesions in multiple sclerosis," (in eng), *Mult Scler,* vol. 14, no. 9, pp. 1214-1219, 2008, doi: 10.1177/1352458508094644.

[150] Y. Chen, E. M. Haacke, and E. Bernitsas, "Imaging of the Spinal Cord in Multiple Sclerosis: Past, Present, Future," (in eng), *Brain Sci,* vol. 10, no. 11, p. 857, 2020, doi: 10.3390/brainsci10110857.

[151] J. Du *et al.*, "Ultrashort echo time (UTE) magnetic resonance imaging of the short T2 components in white matter of the brain using a clinical 3T scanner," (in eng), *Neuroimage,* vol. 87, pp. 32-41, 2014, doi: 10.1016/j.neuroimage.2013.10.053.

[152] Y. J. Ma *et al.*, "Ultrashort echo time (UTE) magnetic resonance imaging of myelin: technical developments and challenges," *Quant Imaging Med Surg,* vol. 10, no. 6, pp. 1186-1203, Jun 2020, doi: 10.21037/qims-20-541.

[153] C. Guglielmetti, T. Boucneau, P. Cao, A. Van der Linden, P. E. Z. Larson, and M. M. Chaumeil, "Longitudinal evaluation of demyelinated lesions in a multiple sclerosis model using ultrashort echo time magnetization transfer (UTE-MT) imaging," (in eng), *Neuroimage,* vol. 208, pp. 116415-116415, 2020, doi: 10.1016/j.neuroimage.2019.116415.

[154] Y. Ma *et al.*, "Three-Dimensional Inversion Recovery Ultrashort Echo Time (3D IR-UTE) Magnetic Resonance Imaging of Myelin in Rats and Mice Subject to Cuprizone Treatment."

[155] J. M. Theysohn *et al.*, "7 tesla MRI of microbleeds and white matter lesions as seen in vascular dementia," *Journal of Magnetic Resonance Imaging,* vol. 33, no. 4, pp. 782-791, 2011/03/29 2011, doi: 10.1002/jmri.22513.

[156] A. M. Afsahi *et al.*, "High-Contrast Lumbar Spinal Bone Imaging Using a 3D Slab-Selective UTE Sequence," *Front Endocrinol (Lausanne),* vol. 12, p. 800398, 2021, doi: 10.3389/fendo.2021.800398.

[157] R. A. Horch, K. Wilkens, D. F. Gochberg, and M. D. Does, "RF coil considerations for short-T2 MRI," (in eng), *Magnetic resonance in medicine,* vol. 64, no. 6, pp. 1652-1657, 2010, doi: 10.1002/mrm.22558.